\documentclass[12pt]{article}
\usepackage{epsfig}
\usepackage{pstricks,rotating, graphicx}
\usepackage{a4}
\usepackage{latexsym}
\usepackage{cite}

\usepackage{color}
\usepackage{colordvi}
\usepackage[hang]{subfigure}

\graphicspath{{../Pics/}}
\DeclareGraphicsExtensions{.eps,.ps}

\usepackage[width=16.5cm,left=2.2cm,top=1.75cm,height=24.cm]{geometry}

\usepackage{pslatex}
\usepackage{txfonts}
\usepackage[latin1]{inputenc}
\usepackage[T1]{fontenc}

\newcommand{\msbar}{$\overline{\mathrm{MS}}\, $}

\begin{document}

\begin{titlepage}
\noindent
DESY 13-183\\
DO-TH 13/26\\
LPN 13-068\\
SFB/CPP-13-71\\
October 2013 \\
\vspace{1.75cm}

\begin{center}
  {\bf
    \Large
    The ABM parton distributions tuned to LHC data \\
  }
  \vspace{1.5cm}
  {\large
    S.~Alekhin$^{\, a,b,}$\footnote{{\bf e-mail}: sergey.alekhin@ihep.ru},
    J.~Bl\"umlein$^{\, a,}$\footnote{{\bf e-mail}: johannes.bluemlein@desy.de},
    and S.~Moch$^{\, a,c,}$\footnote{{\bf e-mail}: sven-olaf.moch@desy.de} \\
  }
  \vspace{1.2cm}
  {\it
    $^a$Deutsches Elektronensynchrotron DESY \\
    Platanenallee 6, D--15738 Zeuthen, Germany \\
    \vspace{0.2cm}
    $^b$Institute for High Energy Physics \\
    142281 Protvino, Moscow region, Russia\\
    \vspace{0.2cm}
    $^c$ II. Institut f\"ur Theoretische Physik, Universit\"at Hamburg \\
    Luruper Chaussee 149, D--22761 Hamburg, Germany \\
  }
  \vspace{1.4cm}
  \large {\bf Abstract}
  \vspace{-0.2cm}
\end{center}
We present a global fit of parton distributions at next-to-next-to-leading order (NNLO) in QCD. 
The fit is based on the world data for deep-inelastic scattering, 
fixed-target data for the Drell-Yan process and includes, for the first time, 
data from the Large Hadron Collider (LHC) for the Drell-Yan process and the
hadro-production of top-quark pairs.
The analysis applies the fixed-flavor number scheme for $n_f=3,4,5$, 
uses the \msbar\ scheme for the strong coupling $\alpha_s$ and the heavy-quark masses 
and keeps full account of the correlations among all non-perturbative parameters.
At NNLO this returns the values of $\alpha_s(M_Z) = 0.1132 \pm 0.0011$ and 
$m_t({\rm pole}) = 171.2 \pm 2.4$~GeV for the top-quark pole mass.
The fit results are used to compute benchmark cross sections 
for Higgs production at the LHC to NNLO accuracy. 
We compare our results to those obtained by other groups and show that differences 
can be linked to different theoretical descriptions of the underlying physical processes.
\end{titlepage}

\newpage
\setcounter{footnote}{0}

\section{Introduction}
\label{sec:intro}

Our knowledge of the proton structure builds on the accumulated world data 
from the deep-inelastic scattering (DIS) experiments, which cover a broad kinematic range 
in terms of the scaling variable $x$ and the momentum $Q^2$ transferred to the proton~\cite{Beringer:1900zz}.
These data have been gathered in a variety of different scattering experiments, 
either on fixed targets or through colliding beams, and in the past two decades, 
especially the HERA electron-proton collider has contributed significantly 
with very accurate measurements spanning a wide range in $x$ and $Q^2$.
Thus, DIS world data form the backbone for the determination of the parton distribution functions
(PDFs) in the QCD improved parton model.

Modern PDFs, however, are expected to provide an accurate description of the
parton content of the proton not only in a kinematic region for $x$ and $Q^2$ as
wide as possible, but to deliver also information on the flavor composition of
the proton as well as on other non-perturbative parameters associated to the
observables under consideration, such as the strong coupling constant $\alpha_s$
or the masses of the heavy quarks charm, bottom and top.
In the theoretical predictions the values for these quantities are often 
correlated with the PDFs and, therefore, have to be determined simultaneously 
in a fit.

A comprehensive picture of a composite object such as the proton 
does not emerge without the need for additional assumptions by relying, 
e.g., on DIS data from the HERA collider alone.
Therefore, global PDF fits have to include larger sets of precision data for different processes,
which have to be compatible, though.
The release of the new data for so-called standard candle processes, i.e., 
precisely measured and theoretically well-understood Standard Model (SM) scattering
reactions, initiates three steps in the analysis:
\begin{itemize}
\item[i)] check of compatibility of the new data set with the available world data
\item[ii)] study of potential constraints due to the addition of the new data set to the fit
\item[iii)] perform a high precision determination of the non-perturbative
  parameters: PDFs, $\alpha_s(M_Z)$ and heavy-quark masses.
\end{itemize}
Of course, at every step QCD precision analyses have to provide a detailed account of the systematic errors and 
have to incorporate all known theoretical corrections.
At the Large Hadron Collider (LHC) PDFs are an indispensable 
ingredient in almost every experimental analysis and the publication of data 
for $W^\pm$- or $Z$-boson, top-quark pair or jet-production from the runs 
at $\sqrt{s}=7$ and $8$~TeV center-of-mass (c.m.s.) energy 
motivates the investigation of potential constraints on SM parameters anew.

Precision data, of course, has to be confronted to high precision theory descriptions. 
In a hadron collider environment, the reduction of the theoretical uncertainty
below ${\cal O}(10\%)$ cannot be achieved without recourse 
to predictions at next-to-next-to-leading order (NNLO) in QCD~\cite{Moch:2004pa,Vogt:2004mw}
which has thus become the standard paradigm of QCD precision analyses  
of the proton's parton content~\cite{Alekhin:2001ih}.
The PDF fits ABKM09~\cite{Alekhin:2009ni} and, subsequently, ABM11~\cite{Alekhin:2012ig} 
on which the current analysis is building, have been performed precisely in this spirit.
At the same time, the NNLO paradigm has motivated continuous improvements in the theory description 
of processes where only next-to-leading order (NLO) corrections are available, such as the hadro-production 
of jets.

In the current article, we are, for the first time, tuning the ABM PDFs to
the available LHC data for a number of standard candle processes including 
$W^\pm$- and $Z$-boson production as well as $t{\bar t}$-production.
We are demonstrating overall very good consistency of the ABM11 PDFs with the available LHC data.
Particular aspects of these findings have been reported 
previously~\cite{Alekhin:2010dd,Alekhin:2013dmy,Alekhin:2013kla,Alekhin:2013qqa,Alekhin:2013nua}.
Subsequently, we perform a global fit to obtain a new ABM12 PDF set and we discuss in detail
the obtained results for the PDFs, $\alpha_s(M_Z)$ and the quark masses
along with their correlations and the goodness of fit.

The outline of the article is as follows.
We recall in Sec.~\ref{sec:basics} the footing of our fit and present the
basic improvements in the theory description and the new data sets included.
These encompass the charm-production and high-$Q$ neutral-current HERA data discussed 
in Sections.~\ref{sec:herac} and \ref{sec:heraq2}, the $W^\pm$- and $Z$-boson production data
from the LHC investigated in Section~\ref{sec:dy} and, likewise, 
in Sec.~\ref{sec:ttbar} data for the total cross section of $t{\bar t}$-production.
The results for ABM12 PDFs are discussed in Section~\ref{sec:results} in a detailed 
comparison with the ABM11 fit in Sec.~\ref{sec:comp} 
and with emphasis on the strong coupling constant and the charm quark mass, cf. Section~\ref{sec:const}.
Finally, in Section~\ref{sec:standard-candles} we provide cross section
predictions of the ABM12 PDFs for a number of standard candle processes 
and the dominant SM Higgs production channel.
Appendix~\ref{sec:appA} describes a fast algorithm for dealing with those 
iterated theoretical computations in the PDF fit, which are very time-consuming.

\renewcommand{\theequation}{\thesection.\arabic{equation}}
\setcounter{equation}{0}
\renewcommand{\thefigure}{\thesection.\arabic{figure}}
\setcounter{figure}{0}
\renewcommand{\thetable}{\thesection.\arabic{table}}
\setcounter{table}{0}
\section{New data included and the theory update}
\label{sec:basics}                      

The present analysis is an extension of the earlier ABM11 
fit~\cite{Alekhin:2012ig} based on the DIS and DY data 
and performed in the NNLO accuracy. The improvements are 
related to adding recently published data relevant for the 
PDF determination:

\begin{itemize}

\item semi-inclusive charm DIS production data obtained by combination of the 
H1 and ZEUS results~\cite{Abramowicz:1900rp}.
This data set provides an improved constraint on the 
low-$x$ gluon and sea-quark distribution and allows amended validation   
of the $c$-quark production mechanism in the DIS. 

\item the neutral-current DIS inclusive data with the momentum transfer 
$Q^2>1000~{\rm GeV}^2$ obtained by the HERA experiments~\cite{Aaron:2009aa}. 
These data allow to check the 3-flavor scheme used in our analysis 
up to very high momentum transfers and, besides,
to improve somewhat the determination of the quark distributions at $x \sim 0.1$. 

\item the DY data obtained by the LHC 
experiments~\cite{Aad:2011dm,Aaij:2012vn,Chatrchyan:2012xt,Aaij:2012mda}
improve the determination of the quark distribution at $x \sim 0.1$, and 
in particular provide a constraint on the $d$-quark distribution, which is 
not sensitive to the correction on the nuclear effects in deuteron.  

\item the total top-quark pair-production cross section data from 
LHC~\cite{ATLAS:2012fja,Chatrchyan:x2012bra,ATLAS:2012jyc,CMS:2012dya,CMS:2012gza} 
and the Tevatron combination~\cite{tevewwg:2012} 
provide the possibility for a consistent determination the top-quark mass 
with full account of the correlations with the gluon PDF and the strong coupling $\alpha_s$. 

\end{itemize}

The theoretical framework of the analysis is properly improved as compared to
the ABM11 fit in accordance with the new data included. In this  
Section we describe details of these improvements related to each 
of the processes and the data sets involved, check agreement of the new data 
with the ABM11 fit, and discuss their impact and the goodness of fit.

\subsection{The HERA charm data}
\label{sec:herac}

The HERA data on the $c$-quark DIS production~\cite{Abramowicz:1900rp}
are obtained by combination of the earlier H1 and ZEUS results. The combined
data span the region of $Q^2=2.5\div 2000~{\rm GeV}^2$ and  
$x=3\cdot 10^{-5}\div 0.05$. The dominating channel of the $c$-quark production at this kinematics 
is the photon-gluon fusion. 
Therefore it provides an additional constraint 
on the small-$x$ gluon distribution. 
Our theoretical description of the HERA data on charm-production is based on the 
fixed-flavor-number (FFN) factorization scheme with 3 light quarks in the 
initial state and the heavy-quarks appearing in the final state. The 
3--flavor Wilson coefficients for the heavy-quark electro-production 
are calculated in NLO~\cite{Laenen:1992xs,Bierenbaum:2009zt} and approximate NNLO corrections
have been also derived recently~\cite{Kawamura:2012cr}. 
The latter are obtained as a combination of the 
threshold resummation calculation~\cite{Presti:2010pd}
and the high-energy asymptotics~\cite{Catani:1990eg} with the
available Mellin moments of the massive operator matrix elements 
(OMEs)~\cite{Bierenbaum:2007qe,Bierenbaum:2008yu,Bierenbaum:2009mv,Ablinger:2010ty}, which provide
matching of these two. 
Two options of the NNLO Wilson coefficient's shape, A and B, given in 
Ref.~\cite{Kawamura:2012cr} encode the remaining uncertainty due to higher 
Mellin moments than given in \cite{Bierenbaum:2009mv}.
In the present analysis, the NNLO corrections are modeled
as a linear combination of the option A and B of 
Ref.~\cite{Kawamura:2012cr} with the interpolation 
parameter $d_N$ with the values of $d_N=0,1$ for the options A and B,
respectively. The interpolation parameter is
fitted to the data simultaneously with other fit parameters
and the shape of the massive NNLO correction preferred by the data
is found to be close to option A with the best fit value of $d_N=-0.10\pm 0.15$.
The same approach was also used in our earlier determination of the 
$c$-quark mass from the DIS data including the HERA charm-production 
ones~\cite{Alekhin:2012vu} with a similar value of $d_N$ obtained.
In our analysis we also employ the running-mass definition for the DIS structure 
functions~\cite{Alekhin:2011jq}.
For comparison, the ABM11 fit is based on the massive NNLO corrections 
stemming from the threshold resummation only~\cite{Presti:2010pd} and 
their uncertainty is not considered. 
\begin{figure}[th!]
\centerline{
  \includegraphics[width=16.0cm]{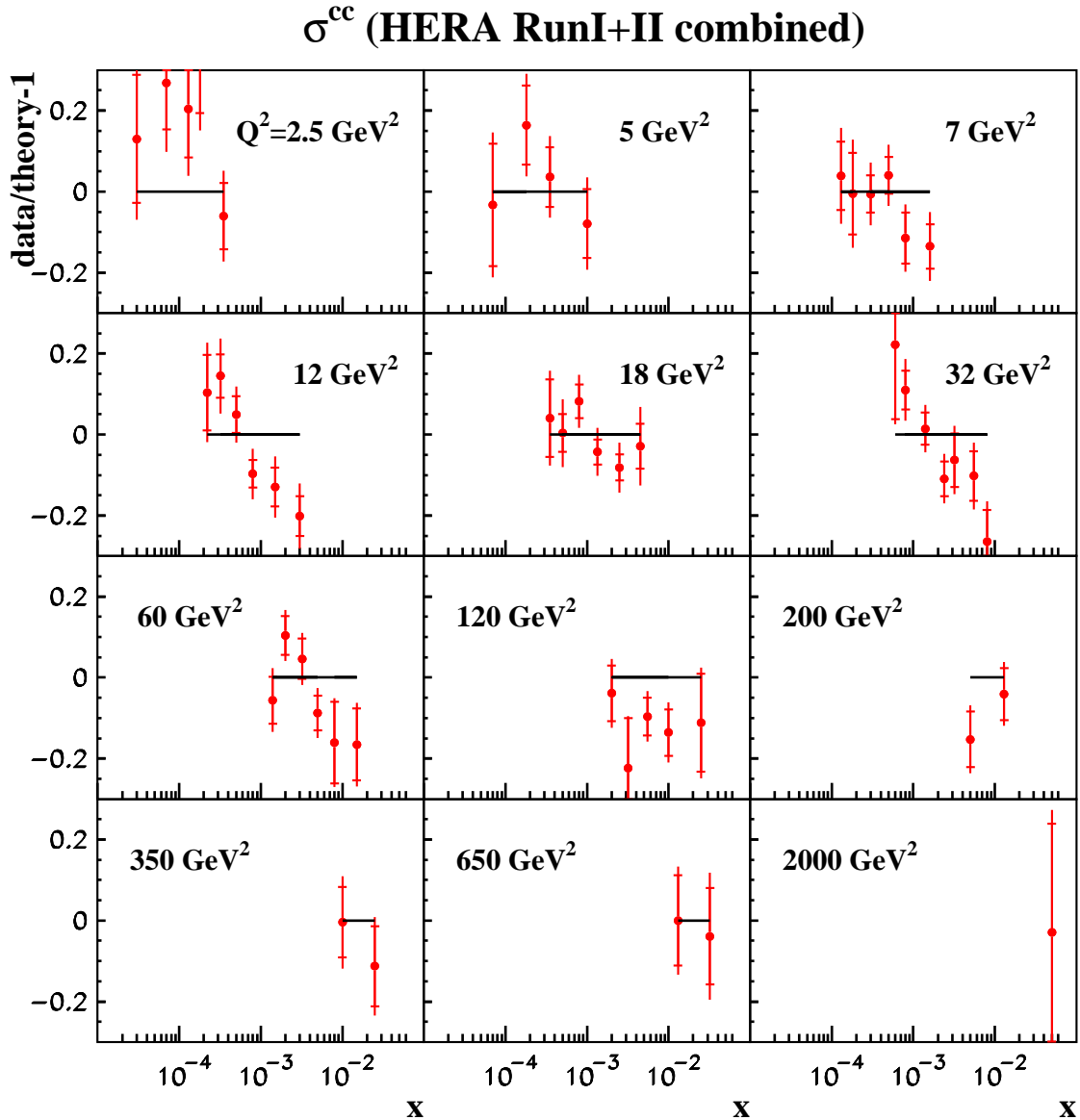}}
  \caption{\small
    \label{fig:herac}
      The pulls versus Bjorken $x$ for
      the HERA combined data on the charm production~\cite{Abramowicz:1900rp} 
      binned in momentum transfer $Q^2$ with respect to our NNLO fit.
}
\end{figure}

The description of the HERA charm data within the ABM12 framework 
is quite good with the value of $\chi^2/NDP=62/52$, where $NDP$ stands for the 
number of data points. The pulls for this data set 
also do not demonstrate any statistically significant trend with respect to either $x$ 
or $Q^2$, cf. Fig.~\ref{fig:herac}. In particular, this gives an argument 
in favor of using the 3-flavor scheme over the full range of existing DIS data 
kinematics. 

\subsection{The high-$Q$ neutral-current HERA data }
\label{sec:heraq2}

\begin{figure}[th!]
\centerline{
  \includegraphics[width=16.0cm]{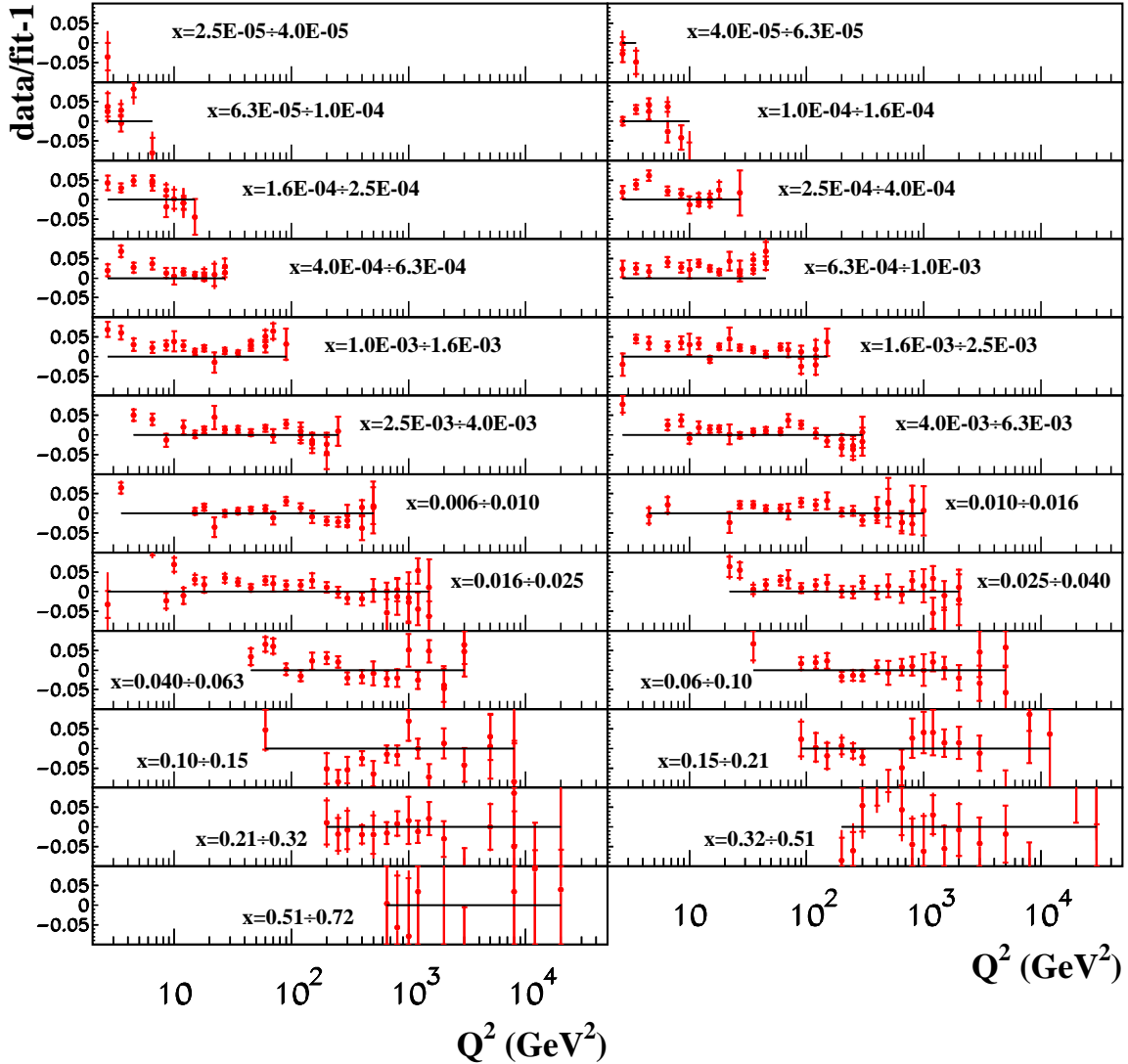}}
  \caption{\small
    \label{fig:heranc}
     The same as in Fig.~\ref{fig:herac} for the  
     pulls of the HERA inclusive combined data~\cite{Aaron:2009aa}  
   binned in Bjorken $x$ versus momentum transfer $Q^2$.
}
\end{figure}

The HERA data for $Q^2>1000~{\rm GeV}^2$ newly added to our analysis are part of the 
combined inclusive sample produced using the H1 and ZEUS statistics
collected during Run-I of the HERA operation~\cite{Aaron:2009aa}. Due to kinematic
constraints of DIS these data are localized at relatively large values of $x$, 
where they have 
limited statistical potential for the PDF constraint as compared to 
the fixed-target DIS data used in our analysis. For this reason 
this piece was not used in the ABM11 fit. In the present 
analysis we fill this gap for the purpose of completeness. 
At large $Q^2$ the DIS cross section gets non-negligible contributions due 
to the $Z$-exchange, in addition to the photon-exchange term sufficient
for the accurate description of the data at $Q^2 \ll M_Z^2$, where 
$M_Z$ is the $Z$-boson mass. The $Z$-boson contribution is taken into 
account using the formalism~\cite{Klein:1983vs,Arbuzov:1995id} with account of the 
correction to the massless Wilson coefficients up to NNLO~\cite{Vermaseren:2005qc}. 
In accordance with~\cite{Klein:1983vs} the contribution due to the photon-$Z$ interference term 
dominates over the one for the pure $Z$-exchange at HERA 
kinematics~\footnote{The version 1.6 of the {\tt OPENQCDRAD} code
used in our analysis to compute the DIS structure functions  
including the contribution due to the $Z$-exchange is publicly available 
online~\cite{OPENQCDRAD}.}. 
The values of $\chi^2/NDP$ obtained in our analysis 
for the whole inclusive HERA data set and for its neutral-current 
subset are $694/608$ and $629/540$, respectively. The data demonstrate 
no statistically significant trend with respect to the fit up to the highest 
values of $Q^2$ covered by the data. 
This is illustrated in Fig.~\ref{fig:heranc} with the example of the 
neutral-current $e^+p$ HERA data sample, which contains the 
most accurate HERA measurements at large $Q^2$. For the $e^-p$ sample the 
picture is similar and the total value of $\chi^2/NDP$
obtained for the newly added neutral-current 
data with $Q^2>1000~{\rm GeV}^2$ is $147/142$. For comparison, with the 
cuts of $Q^2>100~{\rm GeV}^2$ and $Q^2>10~{\rm GeV}^2$ we get 
for the same sample the values of 
$\chi^2/NDP=311/344$ and 486/469, respectively. In particular this 
says that the FFN 
scheme used in our analysis is quite sufficient for the description 
of the existing HERA data in the whole kinematical range
(cf.~\cite{Gluck:1993dpa,VFN2013} for more details). 

\subsection{The LHC Drell-Yan data}
\label{sec:dy}

Data on the Drell-Yan (DY) process provide a valuable constraint on the
PDFs extracted from a global PDF fit allowing to disentangle the sea and 
valence quark distributions. At the LHC these data are now available in the 
form of the rapidity distributions of charged leptons produced in the 
decays of the $W$-bosons and/or charged-lepton pairs from the 
$Z$-boson decays~\cite{Aad:2011dm,Aaij:2012vn,Chatrchyan:2012xt,Aaij:2012mda}. 
Due to limited detector acceptance and the $W/Z$ 
event selection criteria the LHC data are commonly obtained in a 
restricted phase space with a cut on the lepton transverse momentum
$P_T^l$ imposed. 
Therefore, taking advantage of these data to constrain the PDFs requires fully 
exclusive calculations of the Drell-Yan process. These  
are implemented up to NNLO in two publicly available codes, 
{\tt DYNNLO}~\cite{Catani:2009sm} and {\tt FEWZ}~\cite{Li:2012wna}. 
Benchmarking these codes we found good mutual agreement for the LHC kinematics. 
We note that with the version 1.3 of {\tt DYNNLO} the numerical convergence is achieved faster
than for version 3.1 of {\tt FEWZ}, 
although even in the former case a typical CPU time required for computing 
rapidity distribution with the accuracy better than 1\% is 200 hours for
the Intel model P9700/2.80~GHz. However, {\tt FEWZ} (version 3.1) provides a 
convenient capability to estimate uncertainties in the cross sections due to
the PDFs. 
\begin{figure}[th!]
\centerline{
  \includegraphics[width=16.0cm]{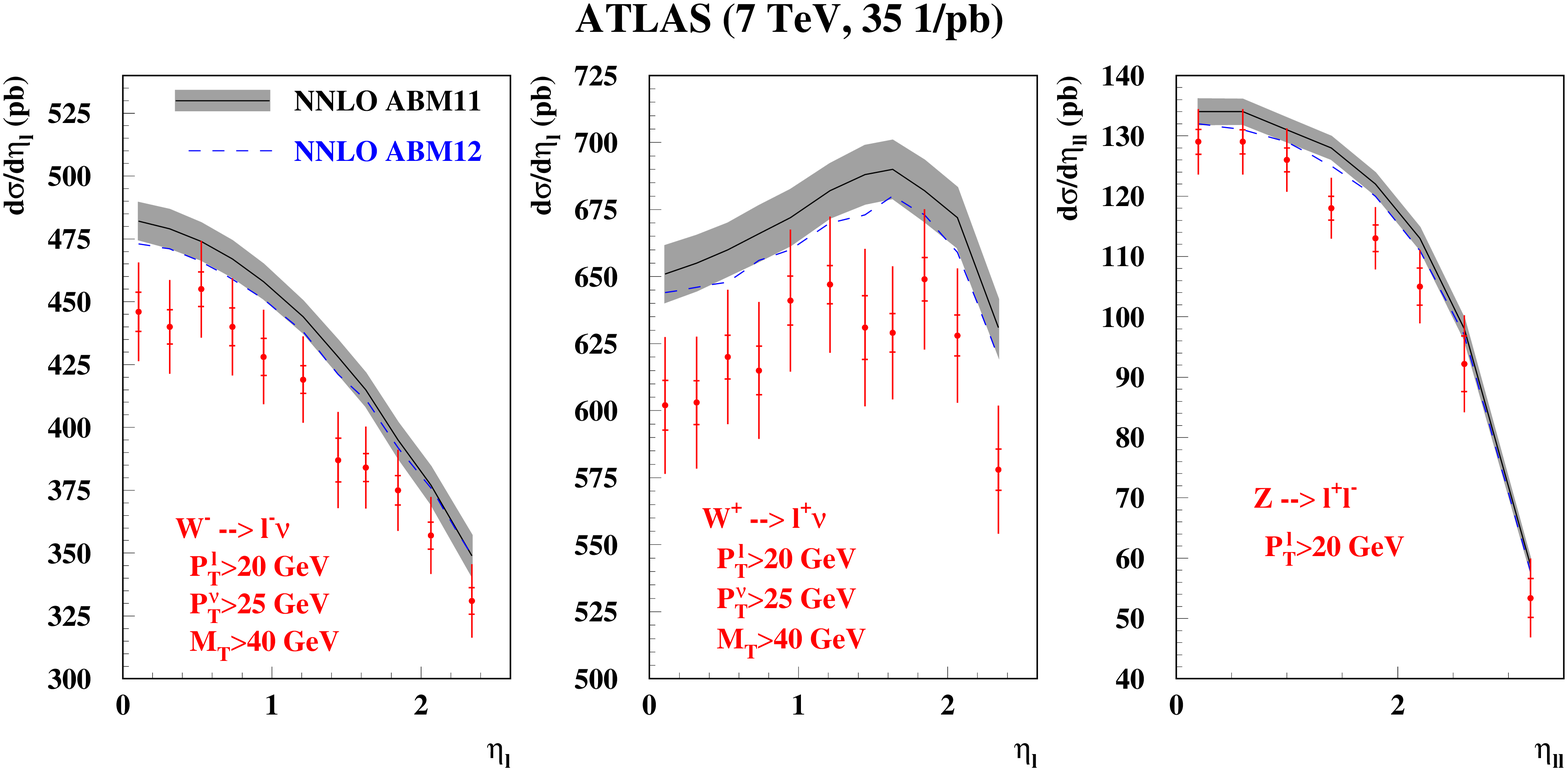}}
  \caption{\small
    \label{fig:atlas}
     The ATLAS data~\cite{Aad:2011dm} on the rapidity distribution 
     of charged leptons produced in the decays of $W^-$- and $W^+$-boson 
     (left and central panel, respectively) and charged lepton pairs 
     from the decays of $Z$-boson (right panel) in comparison with 
     the NNLO calculations based on the ABM11 PDFs (solid curves) taking
     into account the uncertainties due to PDFs (grey area). 
     The dashed curves display the ABM12 predictions. 
     The cuts on the lepton transverse momentum $P_T^l$ and the 
     transverse mass $M_T$ imposed to select a particular process signal are 
     given in the corresponding panels. 
}
\end{figure}
\begin{figure}[th!]
\centerline{
  \includegraphics[width=16.0cm]{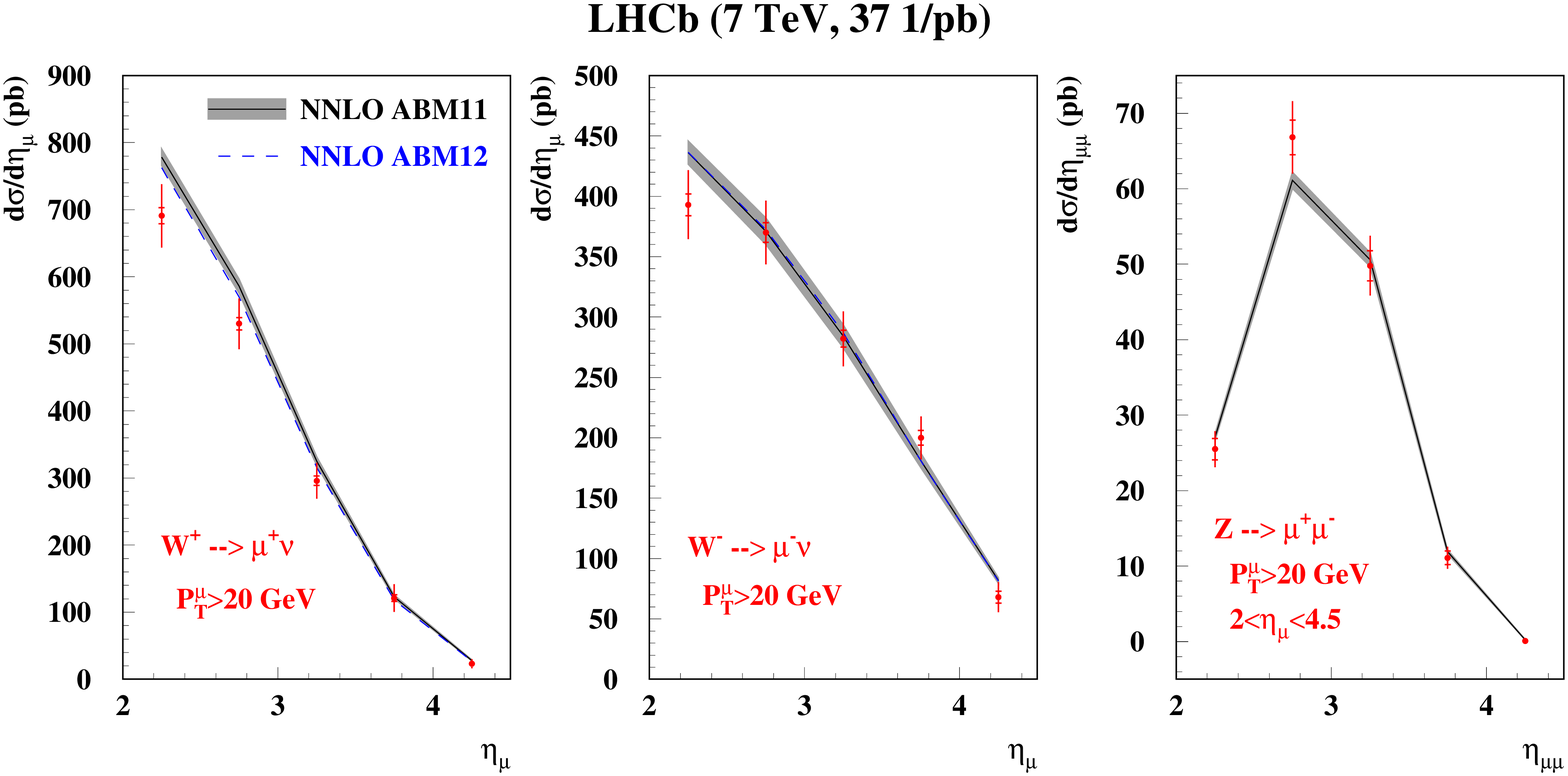}}
  \caption{\small
    \label{fig:lhcb}
    The same as in Fig.~\ref{fig:atlas} for the charged muons 
    rapidity distributions obtained by LHCb~\cite{Aaij:2012vn}.
}
\end{figure}
\begin{figure}[th!]
\centerline{
  \includegraphics[width=8cm]{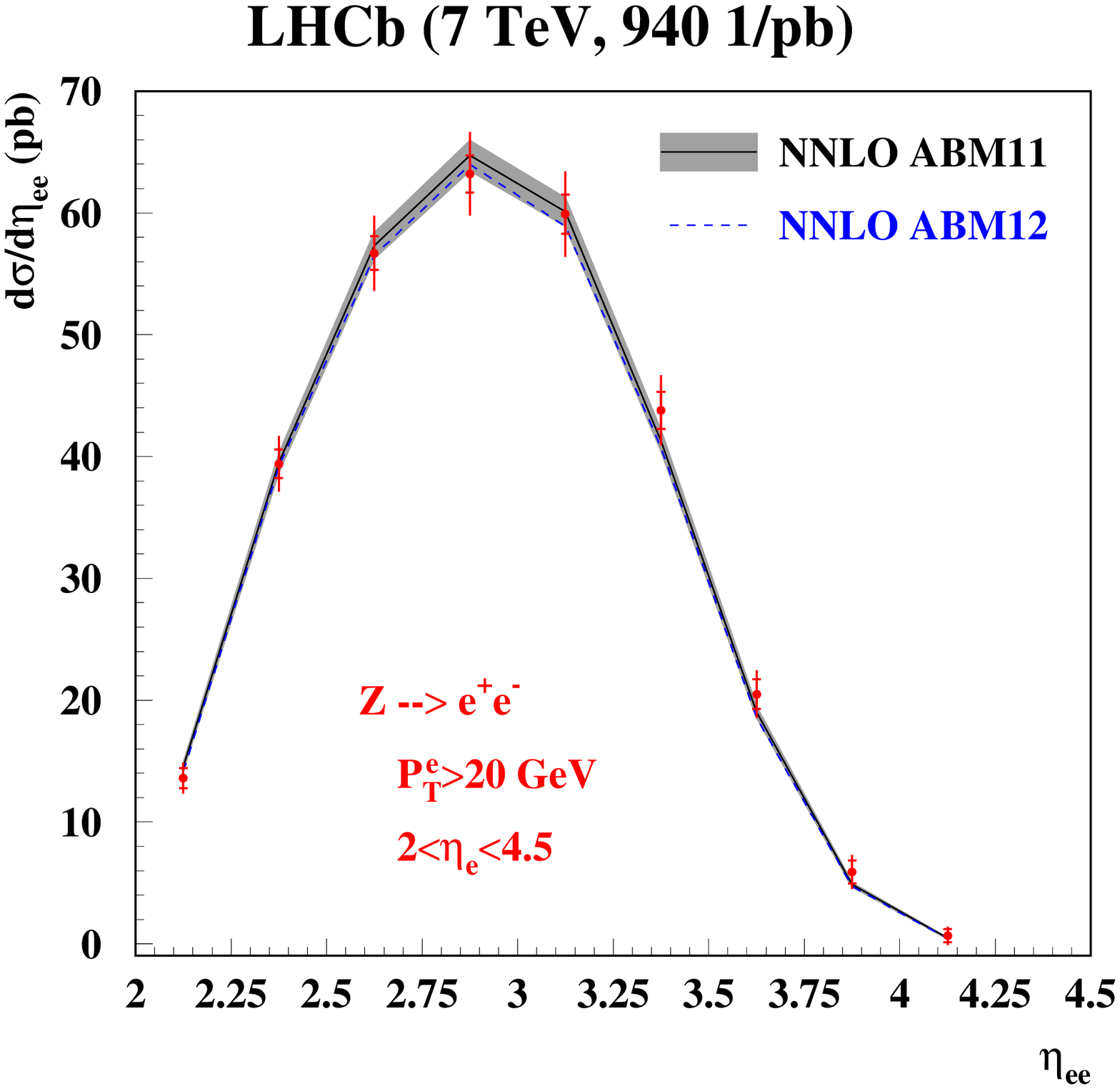}
  \includegraphics[width=8cm]{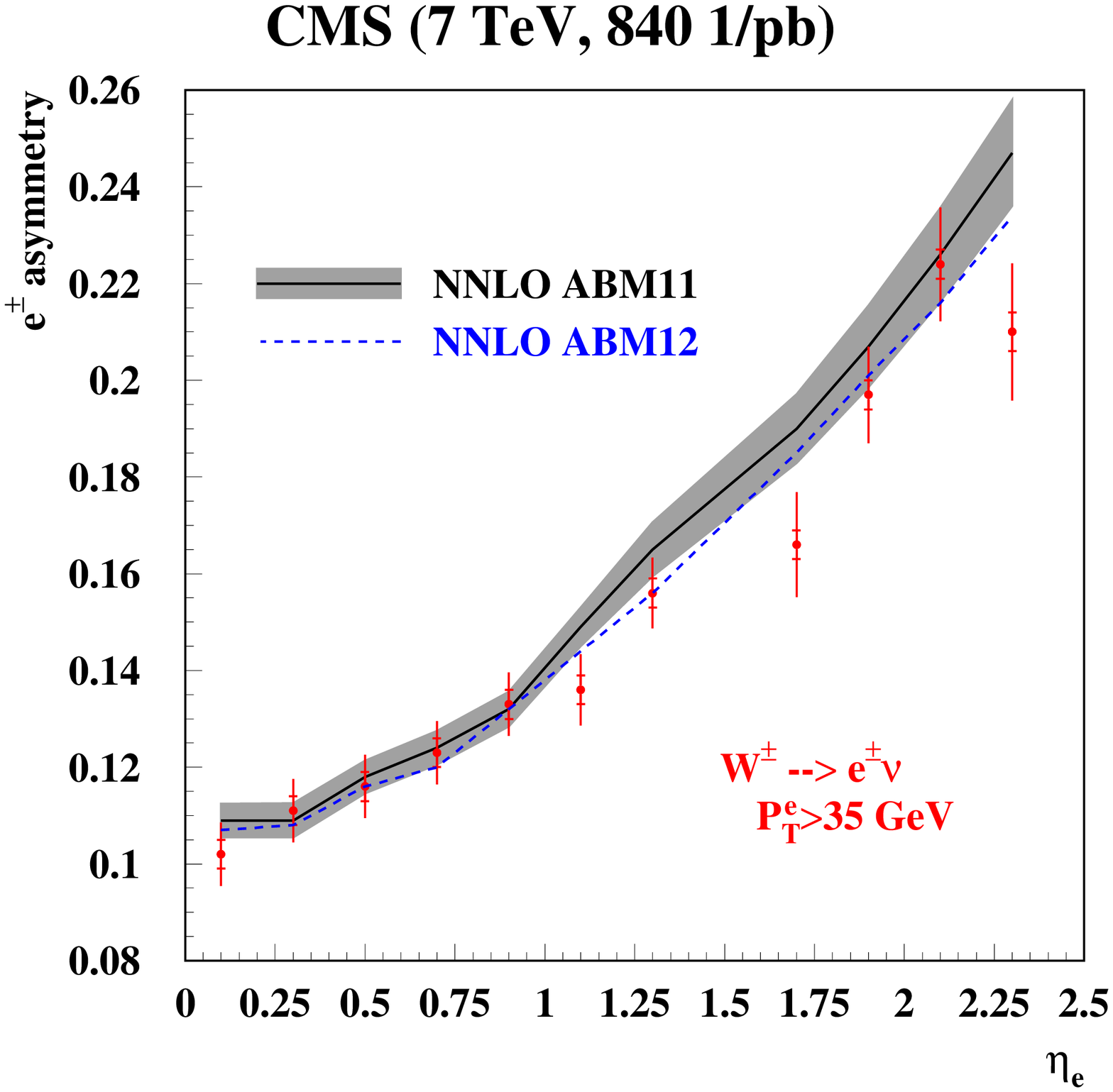}}
  \caption{\small
    \label{fig:lhcbe}
    The same as in Fig.~\ref{fig:atlas} for the LHCb data~\cite{Aaij:2012mda}
    on the rapidity distribution of the $e^+e^-$ pairs produced in the 
    $Z$-boson decays (left panel) and the CMS 
    data~\cite{Chatrchyan:2012xt}
    on the charge asymmetry of electrons produced in the 
    $W^{\pm}$-boson decays (right panel). 
}
\end{figure}
Therefore we use in our analysis the benefits of both codes
combining the central values of {\tt DYNNLO}~(version 1.3) and the 
PDF uncertainties of {\tt FEWZ}~(version 3.1).

The predictions obtained in such a way with the ABM11 
PDFs~\cite{Alekhin:2012ig} are compared to the LHC DY
data~\cite{Aad:2011dm,Aaij:2012vn,Chatrchyan:2012xt,Aaij:2012mda} in 
Figs.~\ref{fig:atlas},~\ref{fig:lhcb} and \ref{fig:lhcbe}.
The predictions systematically overshoot the ATLAS data~\cite{Aad:2011dm}. 
However the offset is within the experimental uncertainty, which is dominated 
by the one of 3.5\% due to the luminosity, cf. Fig~\ref{fig:atlas}.
On the other hand, a good agreement 
is observed for the $Z$-boson data by LHCb~\cite{Aaij:2012mda} in the 
region overlapping with the ATLAS kinematics, cf. Fig~\ref{fig:lhcbe}. 
This signals some 
discrepancy between these two sets of data, which is most likely related to the 
general experimental normalization. 
In any case the normalization off-set cancels in the ratio and the 
ATLAS data on the charged-lepton asymmetry are in a good agreement with our 
predictions~\cite{Aad:2011dm}. This is in some contrast to the CMS 
results where a few data points go lower than the 
ABM11 predictions, cf. Fig~\ref{fig:lhcbe}. 

Agreement between the LHC data and the ABM11 predictions is quantified by 
the following $\chi^2$ functional 
\begin{equation}
\chi^2=\sum_{i,j} (y_i-t_i^{(0)})[C^{-1}]_{ij}(y_j-t_j^{(0)}),
\label{eq:chi}
\end{equation}
where $y_i$ and $t_i^{(0)}$ stand for the measurements and predictions, 
respectively, and $C_{ij}$ is the covariance matrix 
with the indices $i,j$ running over the points in the data set. 
The covariance matrix is constructed as follows
\begin{equation}
C_{ij}=C_{ij}^{exp} + \sum_{k=1}^{N_{unc}}\Delta t_i^{(k)}\Delta t_j^{(k)},
\label{eq:covm}
\end{equation}
where the first term contains information about the experimental errors 
and their correlations and the second term comprises
the PDF uncertainties in predictions. 
The later are quantified as shifts in the predictions due to the variation 
between the central PDF value and the ones encoding the PDF uncertainties. 
For ABM11 the latter appear primarily 
due to the variation of the fitted PDF parameters and, besides, due to the uncertainty
in the nuclear correction applied to the deuteron DIS data. 
Therefore, the total number of PDF uncertainty members is $N_{unc}=N_p+1$, where
$N_p=27$ is the number of eigenvectors in the space of fitted PDF parameters
(cf. the Appendix for more details). 

The experimental covariance matrix for the ATLAS data~\cite{Aad:2011dm} 
is computed by
\begin{equation}
C_{ij}^{exp}=\delta_{ij}\sigma_i^2 + f^{(0)}_i f^{(0)}_j \sum_{l=1}^{31}s_i^ks_j^k,
\label{eq:covexp}
\end{equation}
where $\sigma_i$ are the statistical errors in the data combined in 
quadrature with the uncorrelated errors. Here $s_i^l$ are the correlated 
systematic uncertainties representing 31 independent sources including 
the normalization, and $\delta_{ij}$ stands for the Kronecker symbol. 
In view of the small
background for the $W$- and $Z$-production signal all systematic errors
are considered as multiplicative. Therefore, 
they are weighted with the theoretical predictions $f^{(0)}_i$.
The experimental covariance matrices 
for the CMS and LHCb data of 
Refs.~\cite{Chatrchyan:2012xt,Aaij:2012vn,Aaij:2012mda} 
are employed directly as published in 
Eq.~(\ref{eq:covm}) after re-weighting them by the theoretical predictions
similarly to Eq.~(\ref{eq:covexp}) with the normalization uncertainty 
taken into account in the same way as for the ATLAS data. 
\begin{table}[th!]
\renewcommand{\arraystretch}{1.3}
\begin{center}                   
{\small                          
\begin{tabular}{|c|c|c|c|c|}   
\hline                           
{Experiment}                      
&ATLAS~\cite{Aad:2011dm}                         
&{CMS~\cite{Chatrchyan:2012xt}}  
&{LHCb~\cite{Aaij:2012vn}}
&{LHCb~\cite{Aaij:2012mda}}                            
\\                                                        
\hline
{Final states}                                                   
& $W^+\rightarrow l^+\nu$
& $W^+\rightarrow e^+\nu$
&$W^+\rightarrow \mu^+\nu$
& $Z\rightarrow e^+e^-$                                                        
\\                                                        
& $W^-\rightarrow l^-\nu$
&$W^-\rightarrow e^-\nu$
&$W^-\rightarrow \mu^-\nu$
&                                                         
\\                                                        
& $Z\rightarrow l^+l^-$
&                                                         
&                                                         
&                                                         
\\                                                        
\hline                                                    
{Luminosity (1/pb)}                      
&35                         
&840  
&{37}
&{940}                            
\\                                                        
\hline                                                    
$NDP$
&30                      
&11  
&10
&9                            
\\                                                        
\hline
 $\chi^2$ (ABM11)
 &$35.7(7.7)$
 &$10.6(4.7)$
 &13.1(4.5)
 &11.3(4.2)
\\
\hline
 $\chi^2$ (ABM12)
 &35.6
 &9.3
 &14.4
 &13.4
\\
\hline
\end{tabular}
}
\caption{\small The value of $\chi^2$ obtained for different samples of 
the Drell-Yan LHC data with the NNLO ABM11 PDFs in comparison 
with the one obtained in the ABM12 fit.
The figures in parenthesis give one standard deviation of 
$\chi^2$ equal to $\sqrt{2NDP}$.}
\label{tab:chi2}
\end{center}
\end{table}

The values of $\chi^2$ computed according to Eq.~(\ref{eq:chi}) for each of the LHC DY data sets 
obtained with the ABM11 PDFs are given in Tab.~\ref{tab:chi2}. 
The description quality is somewhat worse for the ATLAS and LHCb muon data, 
however, in general the agreement between the data and predictions is still good. 
The values of $\chi^2/NDP$ are comparable to 1 within the 
statistical fluctuations in $\chi^2$. 
Therefore, the data can be easily accommodated in the ABM fit. Furthermore, 
in this case the PDF variation is expected to be within the ABM11 PDF
uncertainties. This allows to optimize the computation of the involved 
NNLO Drell-Yan corrections in the fit by extrapolation of 
the grid with the pre-calculated predictions for the ABM11 eigenvector basis 
(cf. App.~\ref{sec:appA} for the details on the implementation of this approach).  
The values of $\chi^2$ obtained for the LHC DY data sets in the ABM12 fit 
are quoted in Tab.~\ref{tab:chi2}. 
In this case the PDF uncertainties are irrelevant since the PDFs have been tuned to the data.
Therefore, they are not included into the second term in the covariance matrix Eq.~(\ref{eq:covm}). 
Despite the difference in the definition, the ABM12 
values of $\chi^2$ for the LHC DY data are in a good agreement with the 
ABM11 ones giving additional evidence for the compatibility of 
these data with the ABM11 PDFs. 

\subsection{The data for $t\bar{t}$ production in the ABM12 fit}
\label{sec:ttbar}

At the LHC $t{\bar t}$-pair production proceeds predominantly through initial gluon-gluon scattering. 
Thus, the total $t{\bar t}$ cross section is sensitive to the gluon distribution at effective $x$ values of 
$\langle x \rangle \simeq 2m_t/\sqrt{s} \simeq 0.04 \dots 0.05$ for the runs at $\sqrt{s}=7$ and $8$~TeV c.m.s. energy, 
a region in $x$ which is well constrained by data from the HERA collider, though.

The available data for the total $t{\bar t}$ cross section 
from ATLAS and CMS at $\sqrt{s}=7$ TeV~\cite{ATLAS:2012fja,Chatrchyan:x2012bra}
and at $\sqrt{s}=8$ TeV~\cite{ATLAS:2012jyc,CMS:2012dya,CMS:2012gza} c.m.s. energy 
display good consistency, although, for the data sets at $\sqrt{s}=7$ TeV 
only within the combined uncertainties.
Generally, the systematic and luminosity uncertainties dominate 
over the small statistical uncertainty and the CMS data~\cite{Chatrchyan:x2012bra,CMS:2012dya,CMS:2012gza}
as well as the result from the Tevatron combination~\cite{tevewwg:2012} 
are accurate to ${\cal O}(5 \%)$ while the ATLAS measurements~\cite{ATLAS:2012fja,ATLAS:2012jyc}
have an error slightly larger than ${\cal O}(10 \%)$.

\begin{figure}[t!]
\centerline{
  \includegraphics[width=8.0cm]{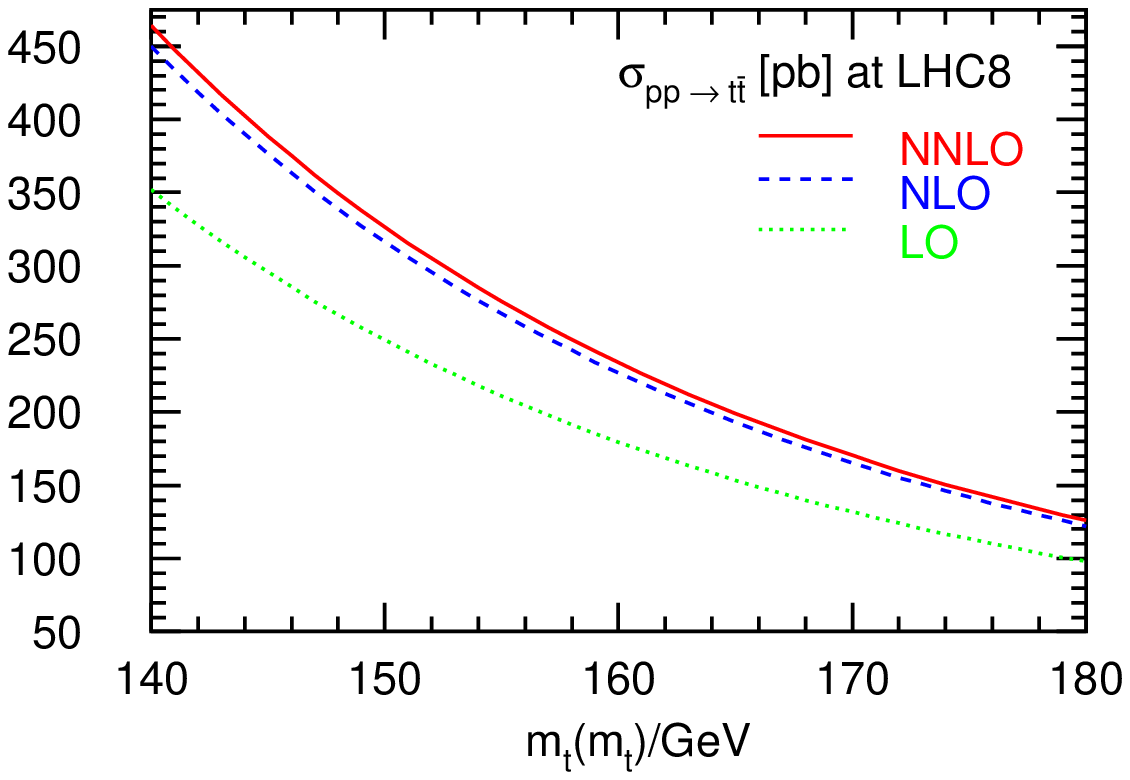}
  \hspace*{5mm}
  \includegraphics[width=8.0cm]{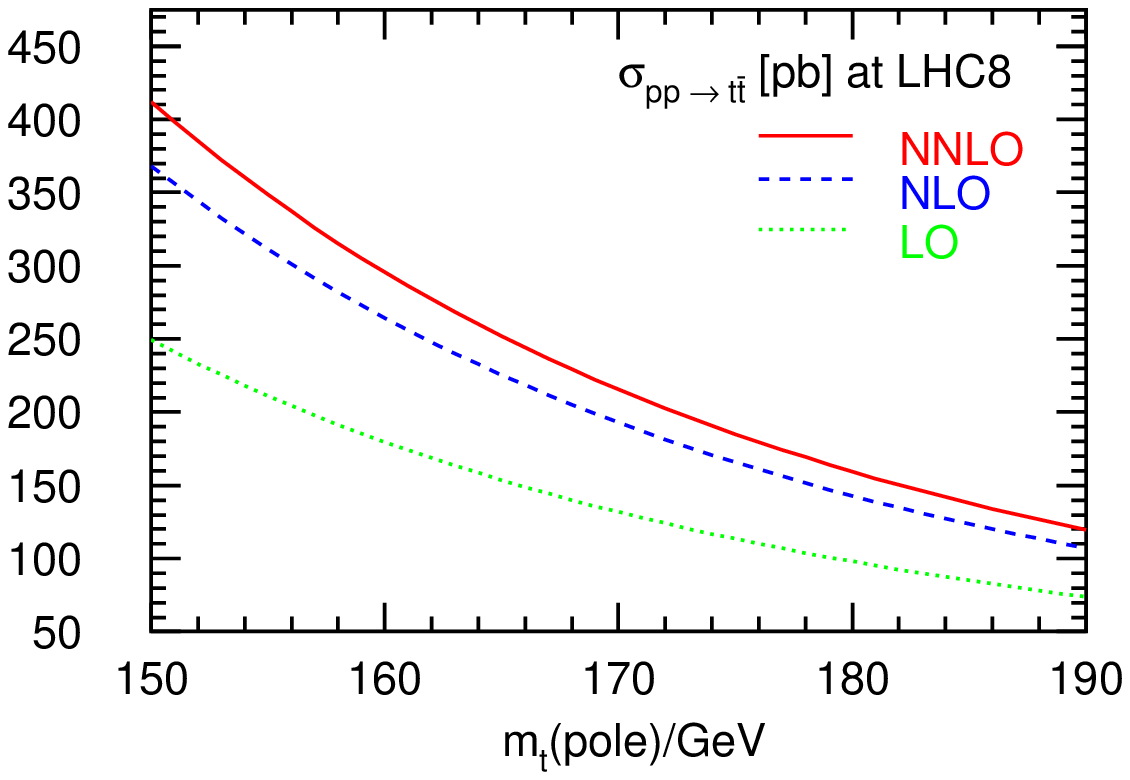}}
  \caption{\small
    \label{fig:ttbar-mass}
    The LO, NLO and NNLO QCD predictions for the 
    $t{\bar t}$ total cross section at the LHC ($\sqrt{s} = 8$~TeV) 
    as a function of the top-quark mass 
    in the \msbar\ scheme $m_t(m_t)$ at the scale $\mu = m_t(m_t)$ (left) 
    and in the on-shell scheme $m_t({\rm pole})$ at the scale $\mu = m_t({\rm pole})$ (right) 
    with the ABM12 PDFs.
  }
\vspace*{10mm}
\centerline{
  \includegraphics[width=8.0cm]{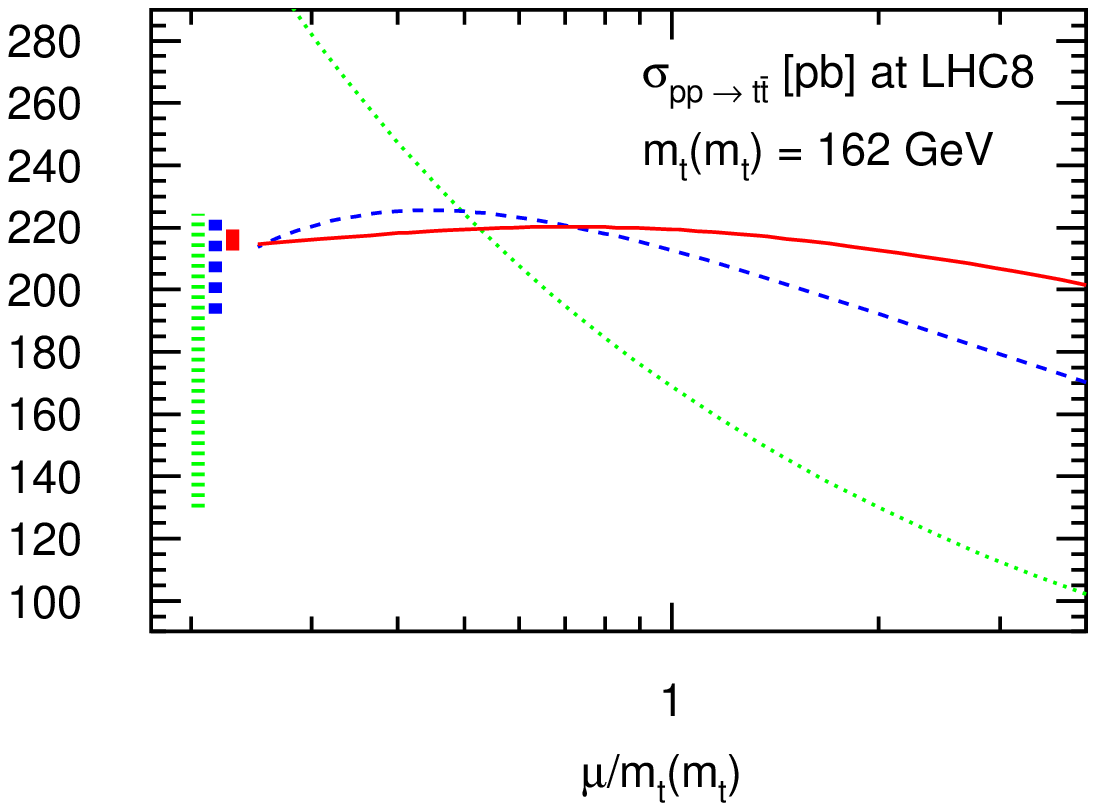}
  \hspace*{5mm}
  \includegraphics[width=8.0cm]{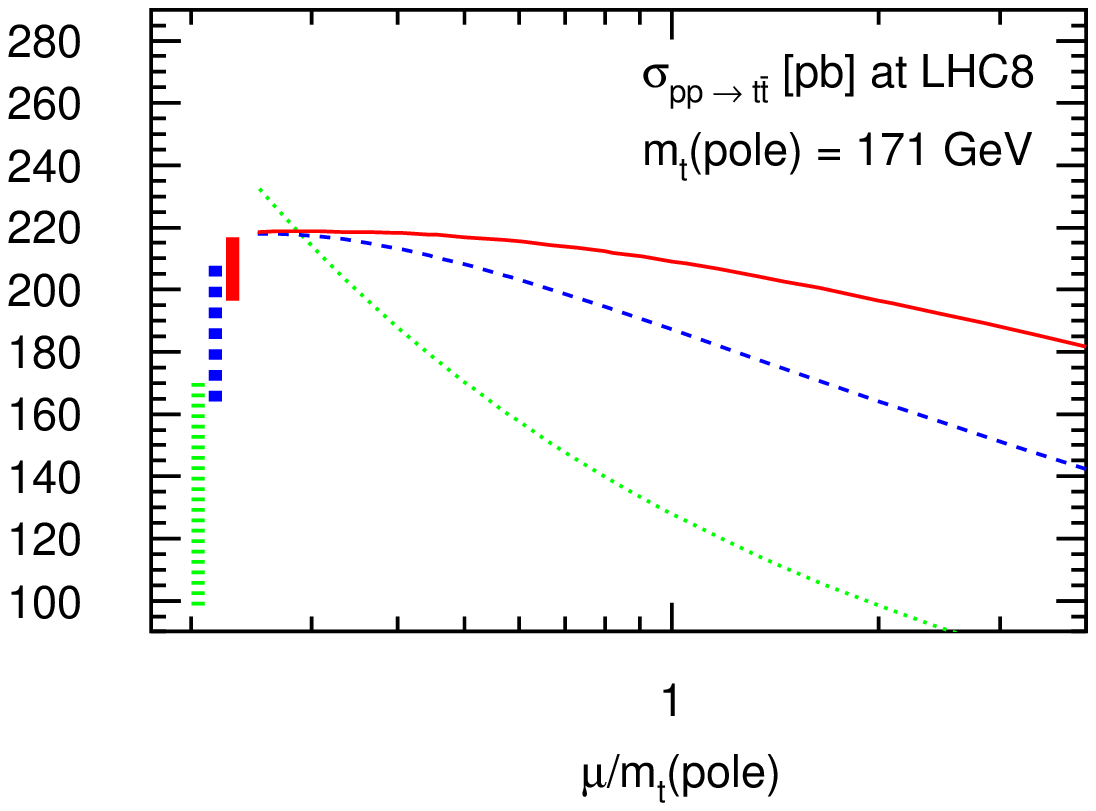}}
  \caption{\small
    \label{fig:ttbar-mu}
    The scale dependence of the LO, NLO and NNLO QCD predictions for the
    $t{\bar t}$ total cross section at the LHC ($\sqrt{s} = 8$~TeV) 
    for a top-quark mass $m_t(m_t)=162$~GeV in the \msbar\ scheme (left) 
    and $m_t({\rm pole})=171$~GeV in the on-shell scheme (right) 
    with the ABM12 PDFs and the choice $\mu = \mu_r = \mu_f$.
    The vertical bars indicate the size of the scale variation in the standard
    range $\mu/m_t({\rm pole}) \in [1/2, 2]$ and $\mu/m_t(m_t) \in [1/2, 2]$, respectively.
}
\end{figure}

The QCD corrections for inclusive $t{\bar t}$-pair production are complete to
NNLO~\cite{Baernreuther:2012ws,Czakon:2012zr,Czakon:2012pz,Czakon:2013goa}, 
so that these data can be consistently added to the ABM11 PDF fit at NNLO.
The theory predictions are available for the top-quark mass 
in the \msbar scheme with $m_t(\mu_r)$ being the running mass~\cite{Langenfeld:2009wd}
as well as for the pole mass $m_t({\rm pole})$ 
in the on-shell renormalization scheme~\cite{Baernreuther:2012ws,Czakon:2012zr,Czakon:2012pz,Czakon:2013goa}. 
The distinction is important, because the theory predictions as a function of the running mass $m_t(\mu_r)$
display much improved convergence and better scale stability of the perturbative expansion~\cite{Langenfeld:2009wd}.
This is illustrated in Figs.~\ref{fig:ttbar-mass} and \ref{fig:ttbar-mu} 
for the total $t{\bar t}$-cross section computed with the program {\tt Hathor} (version 1.5)~\cite{Aliev:2010zk}.
In Fig.~\ref{fig:ttbar-mass} we show the size of the higher order perturbative 
corrections from LO to NNLO taking the PDFs order independent, i.e., the ABM11 set at NNLO,
as a function of the top-quark mass for the LHC at $\sqrt{s}=8$ TeV c.m.s. energy.
Likewise, Fig.~\ref{fig:ttbar-mu} illustrates the scale stability for two
representative top-quark masses, $m_t(m_t)=162$~GeV and $m_t({\rm pole})=171$~GeV.
Figs.~\ref{fig:ttbar-mass} and \ref{fig:ttbar-mu} imply a small residual theoretical uncertainty 
for the $t{\bar t}$-cross section predictions if expressed in terms of the running mass.

\begin{figure}[th!]
\centerline{
  \includegraphics[width=11.75cm]{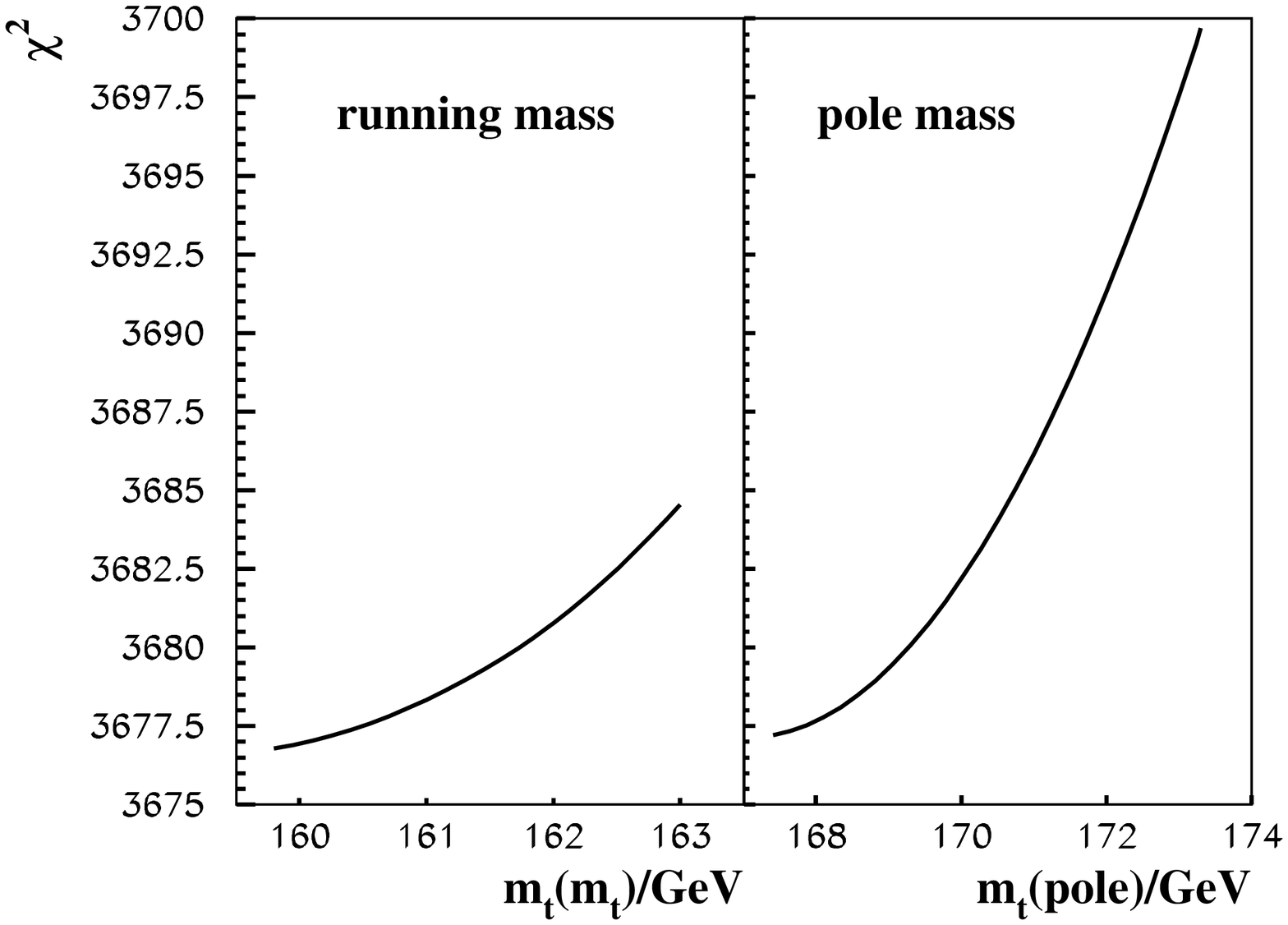}}
  \caption{\small
    \label{fig:chi2}
      The $\chi^2$ profile versus the $t$-quark mass for the variants 
      of ABM12 fit with the $t\bar{t}$ cross section data included
      and different $t$-quark mass definitions: 
      running mass (left) and pole mass (right). 
}
\vspace*{10mm}
\centerline{
  \includegraphics[width=11.75cm]{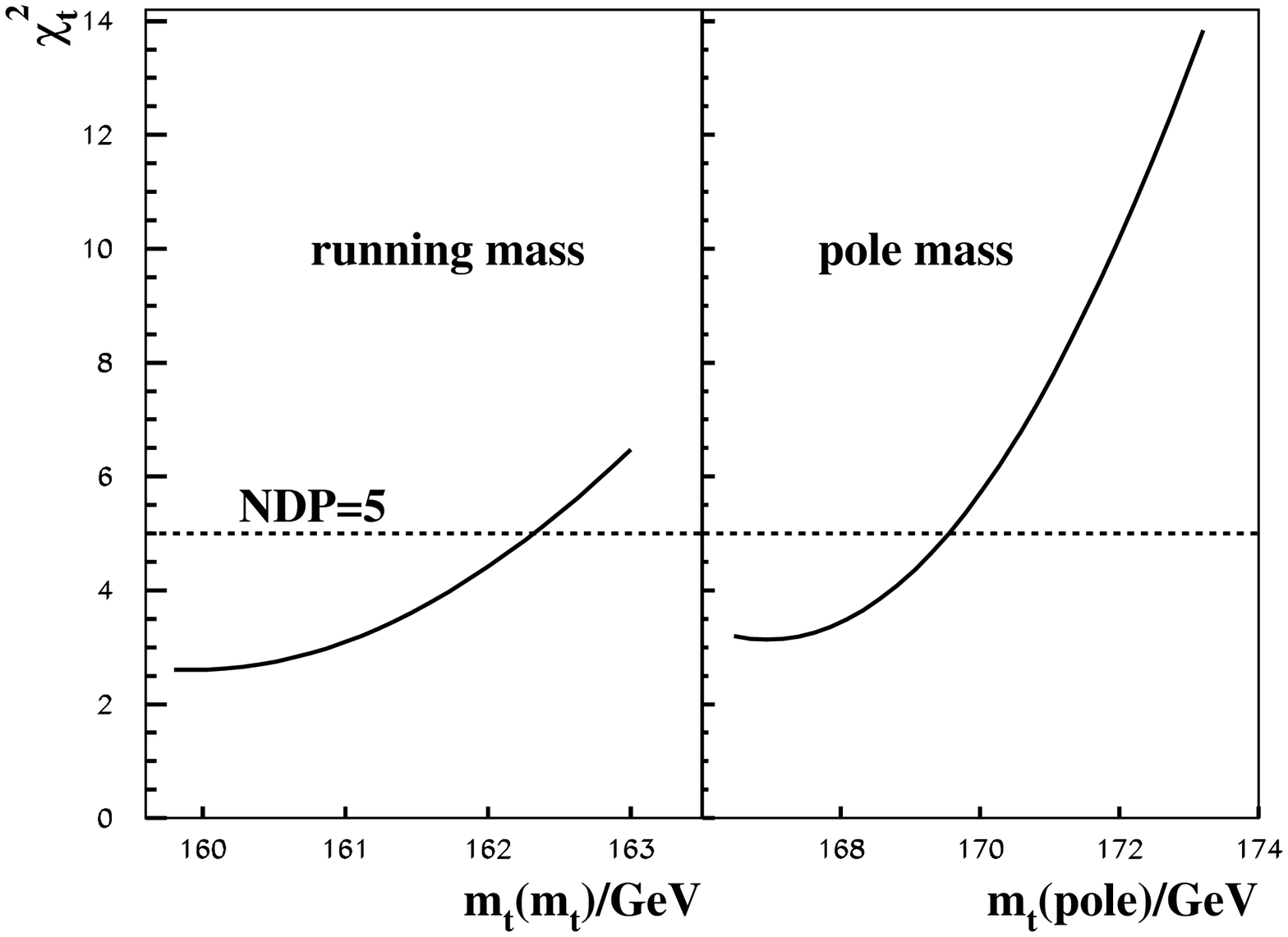}}
  \caption{\small
    \label{fig:chi2t}
     The same as in Fig.~\ref{fig:chi2} for the $t\bar{t}$ cross section data subset. 
     The $NDP=5$ for this subset is displayed by the dashed line. 
}
\end{figure}
\begin{figure}[th!]
\centerline{
  \includegraphics[width=11.75cm]{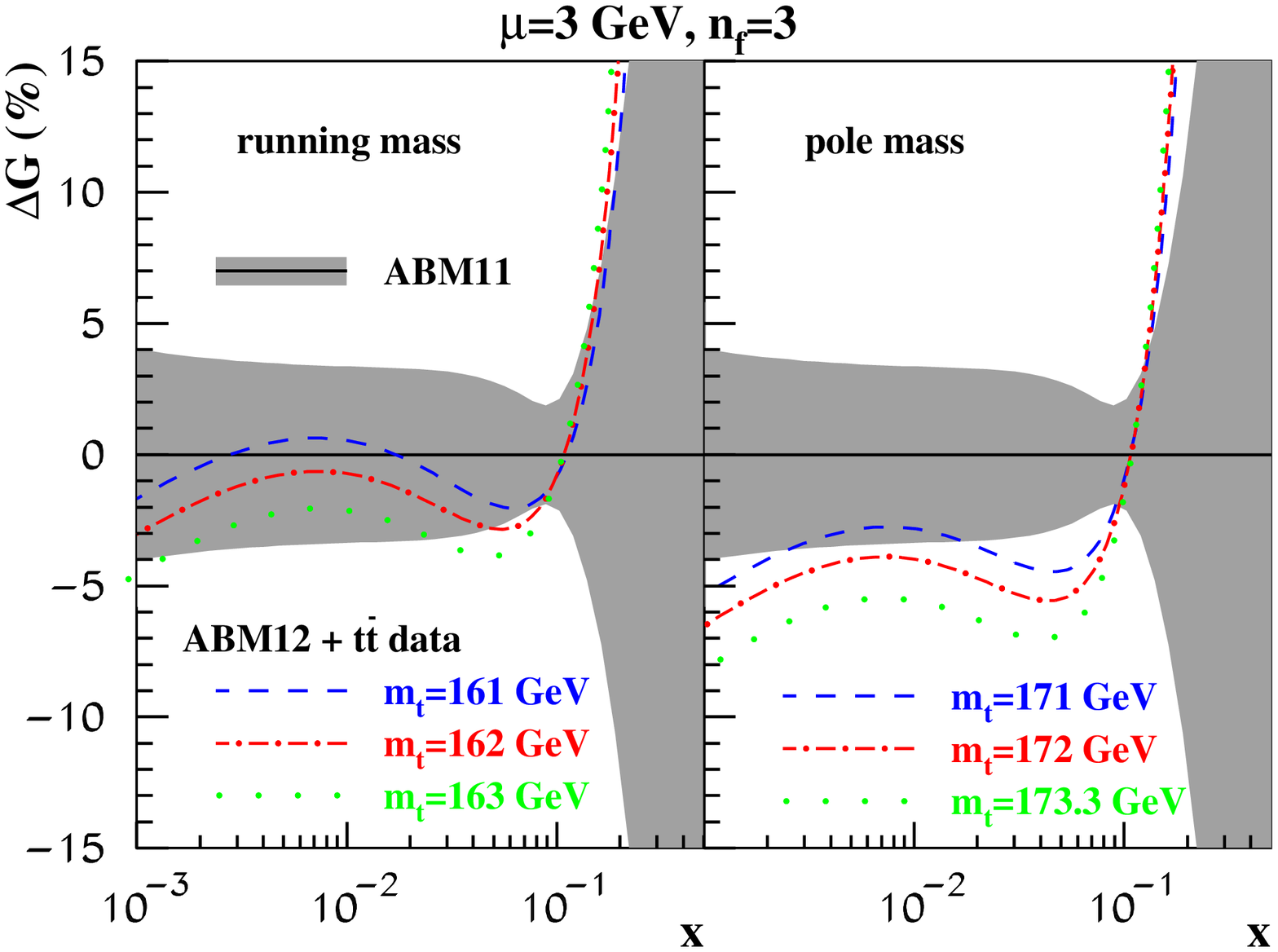}}
  \caption{\small
    \label{fig:pdftq}
     The relative uncertainty in the ABM12 gluon distribution in the 3-flavor
     scheme at the factorization scale of $\mu=3~{\rm GeV}$ (grey area) in comparison 
     to its relative change due to inclusion of the $t\bar{t}$ cross section 
     data with the different mass definitions, 
     running mass (left), pole mass (right),  
     and the $t$-quark mass settings as indicated in the plot.
}
\end{figure}

We have performed a variant of the ABM12 fit, adding the combined $t\bar{t}$ cross section data from LHC and
Tevatron~\cite{ATLAS:2012fja,Chatrchyan:x2012bra,ATLAS:2012jyc,CMS:2012dya,CMS:2012gza,tevewwg:2012} 
to test the impact of these data on the gluon PDF, on the strong coupling $\alpha_s$ 
and on the value and scheme choice for the top-quark mass. 
It is strictly necessary to consider these three parameters together, since they
are strongly correlated in theory predictions for the $t{\bar t}$-cross section at the LHC.
In Figs.~\ref{fig:chi2} and \ref{fig:chi2t} we present the $\chi^2$ profile versus the top-quark mass 
for the variants of the ABM12 fit with the $t\bar{t}$ cross section data included and 
for the two different top-quark mass definitions, 
i.e., the \msbar mass $m_t(m_t)$ and the pole mass $m_t({\rm pole})$.
Fig.~\ref{fig:chi2} displays a steeper $\chi^2$ profile for the pole-mass definition. 
This implies a bigger impact of the $t{\bar t}$-cross section data in the fit
and, as a consequence, greater sensitivity to the theoretical uncertainty at
NNLO and uncalculated higher order corrections to the cross section beyond NNLO.
In contrast, the $\chi^2$ profile for the \msbar mass is markedly flatter.
Fig.~\ref{fig:chi2t} shows the $\chi^2$ profile for the subset of the $t{\bar t}$-cross section data 
with $NDP=5$ and nicely demonstrates that a top-quark mass determination from the fit is feasible. 

If one requires a $\Delta \chi_t^2 = 1$, the value for the \msbar mass is obtained at NNLO
\begin{eqnarray}
\label{eq:mt}
m_t(m_t) = 162.3 \pm 2.3~{\rm GeV}\, ,
\end{eqnarray}
where we define the error in $m_t(m_t)$ due the experimental data, the PDFs and the value of $\alpha_s(M_Z)$
as the difference between the value for $m_t(m_t)$ at $\Delta \chi_t^2 = 1$ 
and the minimum of the $\chi^2$-profile in Fig.~\ref{fig:chi2t}.
The additional theoretical uncertainty from the variation of the factorization
and renormalization scales in the usual range is small, $\Delta m_t(m_t) = \pm 0.7~{\rm GeV}$,
see Fig.~\ref{fig:ttbar-mu} and \cite{Alekhin:2012py}.
Eq.~(\ref{eq:mt}) is equivalent to the top-quark pole mass value of
\begin{eqnarray}
\label{eq:mtpole}
m_t({\rm pole}) \,=\, 171.2 \pm 2.4~{\rm GeV}
\, ,
\end{eqnarray}
using the known perturbative conversion relations~\cite{Gray:1990yh,Chetyrkin:1999qi,Melnikov:2000qh}.
Eq.~(\ref{eq:mtpole}) can be compared to the value of 
$m_t({\rm pole}) \,=\, 169.6 \pm 2.7~{\rm GeV}$ read off from Fig.~\ref{fig:chi2t}.
This indicates good consistency of the procedure and also with the top-quark mass values obtained from other 
determinations\footnote{
The values in Eqs.~(\ref{eq:mt}) and (\ref{eq:mtpole}) supersede the 
top-quark mass determination in \cite{Alekhin:2012py}, 
because full account of the correlations among all non-perturbative parameters is kept.}. 

Having established the sensitivity to the value of the top-quark mass, we have
performed further variants of the ABM12 fit by fixing $m_t(m_t)$ and $m_t({\rm pole})$ 
in order to quantify the impact on the gluon PDF and on $\alpha_s$.
The values for $\alpha_s(M_Z)$ which are obtained in these variants 
span the range $\alpha_s(M_Z)=0.1133 \dots 0.1142$ for $m_t(m_t)=161 \dots 163$~GeV 
and $\alpha_s(M_Z)=0.1144 \dots 0.1154$ for the range of $m_t({\rm pole})=171 \dots 173.3$~GeV. 
The corresponding changes in the gluon PDF are illustrated in Fig.~\ref{fig:pdftq}, which
shows the relative change in the ABM12 gluon distribution at the factorization scale of $\mu=3~{\rm GeV}$
in the 3-flavor scheme due to adding of the $t{\bar t}$-cross section data in the fit 
and fixing $m_t(m_t)$ and $m_t({\rm pole})$ to the values indicated.
For the running-mass definition the changes in the gluon PDF are within the
uncertainties of the nominal ABM12 fit.
In particular, we find $\alpha_s(M_Z)=0.1139(10)$ and a marginal change in the gluon PDF 
for a variant of ABM12 fit with $m_t(m_t)=162$~GeV fixed and with the CMS~\cite{Chatrchyan:x2012bra,CMS:2012dya,CMS:2012gza}
and the Tevatron~\cite{tevewwg:2012} data included, i.e., leaving out the
ATLAS data due to the larger experimental uncertainties. 
This is to be compared with $\alpha_s(M_Z)=0.1133(8)$ for the ABM11 PDF fit
and, again, demonstrates nicely the stability of the analysis, provided all
correlations are accounted for.

We briefly comment here on related studies, that have appeared in the literature.
Ref.~\cite{CMS-PAS-TOP-12-022} determines the strong coupling constant from a
fit to $t{\bar t}$ cross section data and obtains the value of $\alpha_s(M_Z)=0.1185(28)$ for the ABM11 PDFs 
with a fixed $m_t({\rm pole})=173.2$~GeV.
Ref.~\cite{CMS-PAS-TOP-12-022} has used version 1.3 of {\tt  Hathor}~\cite{Aliev:2010zk}, though, 
which returns a slightly different the central value (${\cal O}(1\%)$ change) 
for the cross section compared to version 1.5.
The sensitivity to $\alpha_s$ is determined from fits to sets of PDFs for varying 
values of $\alpha_s(M_Z)$, i.e., using the ABM11 set at NNLO 
({\tt abm11\_5n\_as\_nnlo.LHgrid} in the {\tt LHAPDF} library~\cite{Whalley:2005nh,lhapdf:2013})
which covers the range $\alpha_s=0.105\dots0.12$. 
As a main caveat, the analysis of Ref.~\cite{CMS-PAS-TOP-12-022} misses the
PDFs uncertainties for the PDF sets with varying values of $\alpha_s(M_Z)$
and the correlations of the parameters, i.e., the gluon PDF, $\alpha_s$ and $m_t({\rm pole})$ discussed above.

Ref.~\cite{Czakon:2013tha} explores the constraints on the gluon PDF 
from the same set of LHC and Tevatron $t{\bar t}$ cross section 
data~\cite{ATLAS:2012fja,Chatrchyan:x2012bra,ATLAS:2012jyc,CMS:2012dya,CMS:2012gza,tevewwg:2012} 
considered here. 
The analysis of Ref.~\cite{Czakon:2013tha} uses fixed values for $\alpha_s$ and the pole mass $m_t({\rm pole})$
and, thereby, disregards the correlation of these parameters with the gluon PDF. 
As illustrated in Fig.~\ref{fig:pdftq} this introduces a significant bias so
that the fit results of Ref.~\cite{Czakon:2013tha} are a direct consequence of those assumptions.
Ref.~\cite{Czakon:2013tha} also compares the ABM11 PDFs~\cite{Alekhin:2012ig}
to those data~\cite{tevewwg:2012,ATLAS:2012fja,Chatrchyan:x2012bra,ATLAS:2012jyc,CMS:2012dya,CMS:2012gza}
and quotes a value of $\chi^2=40.2$ for $NDP=5$ (Tab.~7 in Ref.~\cite{Czakon:2013tha}).
Unfortunately, this computation of the $\chi^2$-value is incomplete, 
since it is obtained by neglecting the PDF uncertainties, 
the uncertainty in the value of $m_t({\rm pole})$ as well as other uncertainties, which may have an impact
on the $\chi^2$-value such as the uncertainty in the beam energy, currently estimated to be $1\%$.
The $\chi^2$ profile in Fig.~\ref{fig:chi2t} shows that a faithful account of all sources of
uncertainties and their correlation leads to a very good description of the $t{\bar t}$ cross section data.

\renewcommand{\theequation}{\thesection.\arabic{equation}}
\setcounter{equation}{0}
\renewcommand{\thefigure}{\thesection.\arabic{figure}}
\setcounter{figure}{0}
\renewcommand{\thetable}{\thesection.\arabic{table}}
\setcounter{table}{0}
\section{The ABM12 PDF results}
\label{sec:results}

In this Section 
the results of the ABM12 fit are discussed in detail and compared specifically with the previous ABM11 PDFs.
Regarding the strong coupling constant $\alpha_s(M_Z)$ we also review 
the current situation for $\alpha_s$-determinations from other processes,
where the NNLO accuracy in QCD has been achieved.
Finally, we apply the new ABM12 PDF grids in the format for the {\tt LHAPDF} library~\cite{Whalley:2005nh,lhapdf:2013} 
to compute a number of benchmark cross sections at the LHC.

\begin{figure}[t!]
\centerline{
  \includegraphics[width=16.0cm]{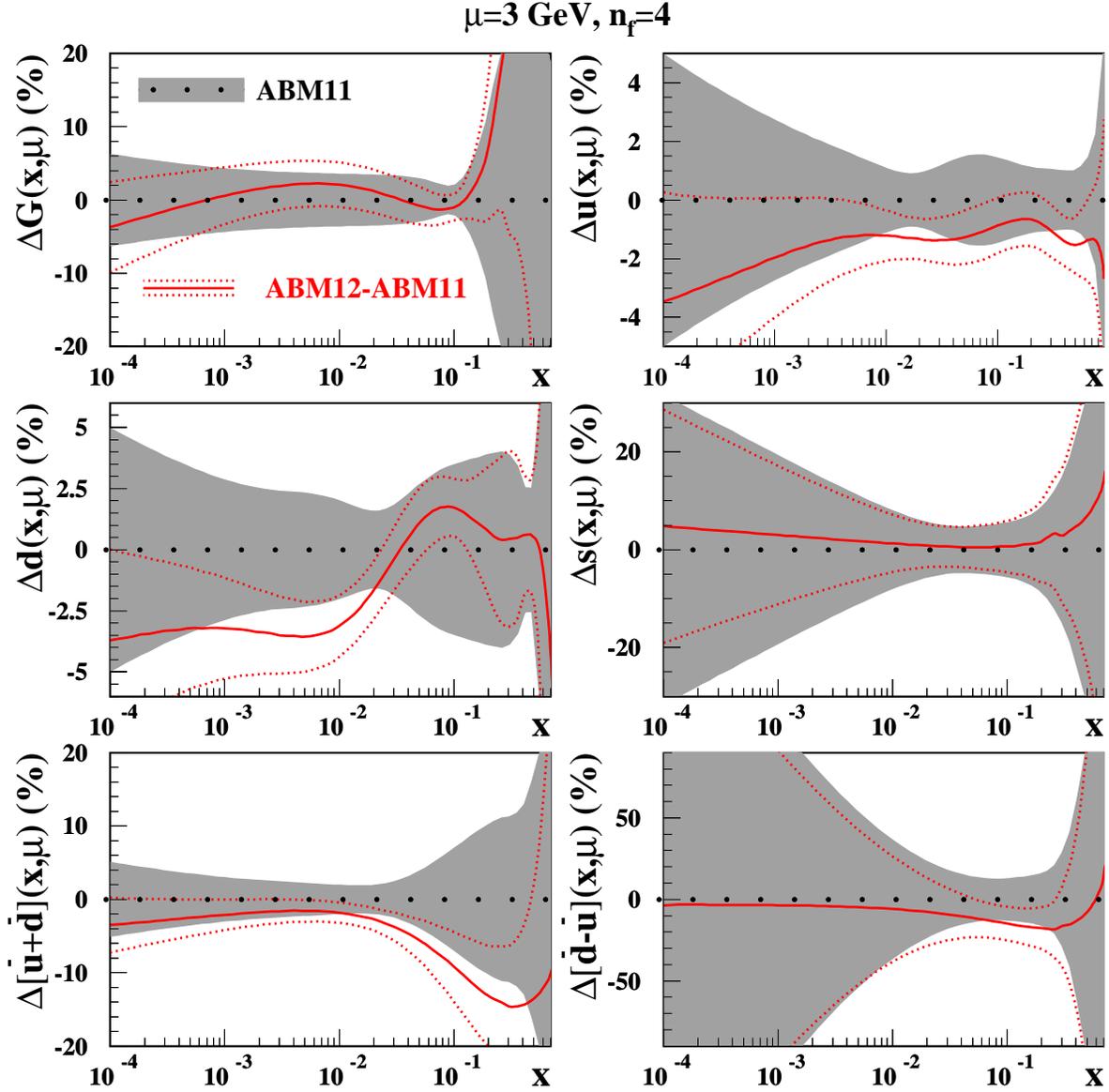}}
  \caption{\small
    \label{fig:abm11}
    The 1$\sigma$ band for the 4-flavor NNLO ABM11 
    PDFs~\cite{Alekhin:2012ig} at the scale 
    of $\mu=3~{\rm GeV}$ versus $x$ (shaded area) 
    compared with the relative difference between ABM11 PDFs and the ABM12 ones 
 obtained in this analysis (solid lines). The dotted lines display 
 $1\sigma$ band for the ABM12 PDFs.
}
\end{figure}
\begin{figure}[th!]
\centerline{
  \includegraphics[width=16.0cm]{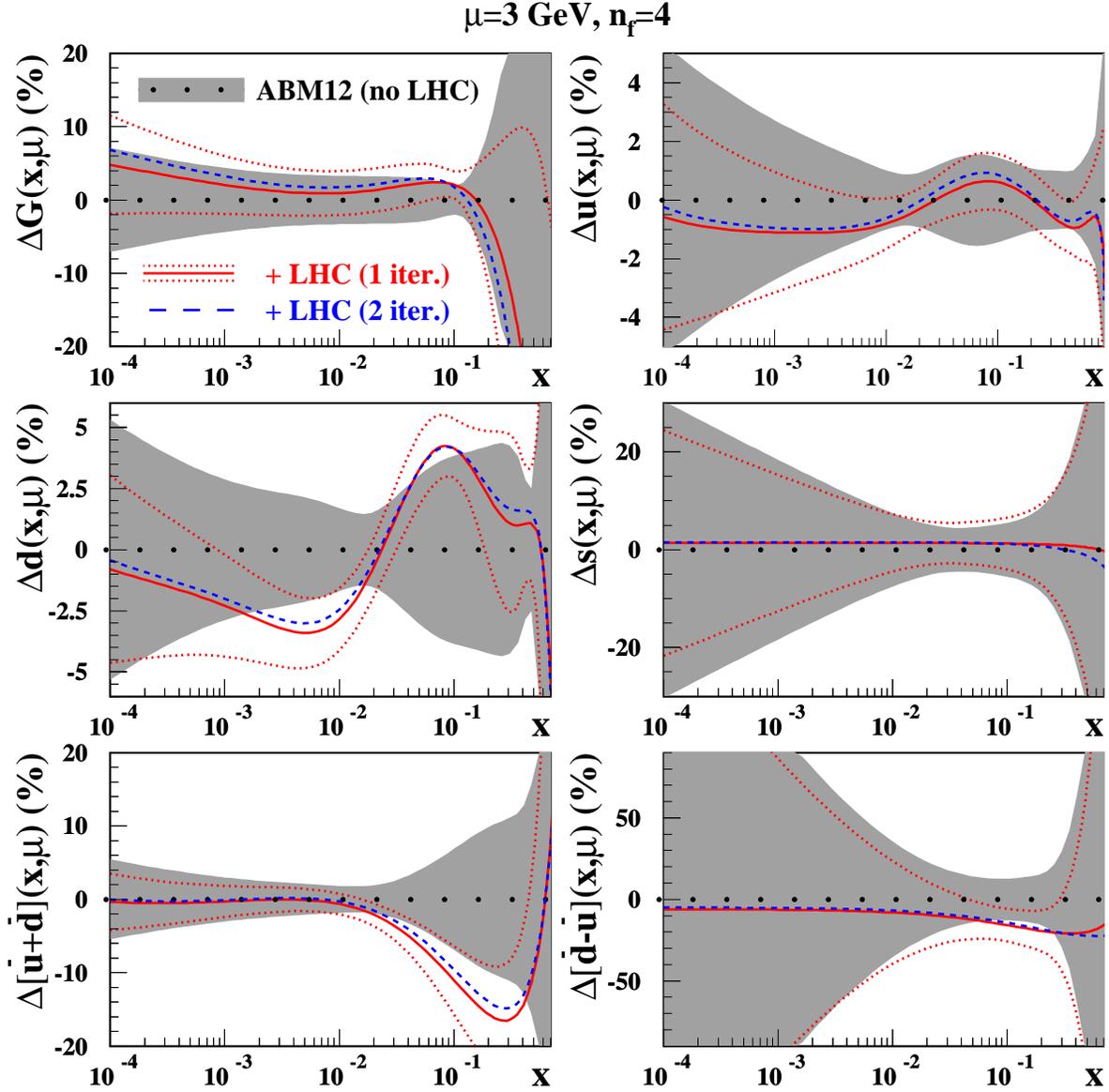}}
  \caption{\small
    \label{fig:abm12}
     The same as in Fig.~\ref{fig:abm11} for the 1$\sigma$ band obtained 
in the variant of the ABM12 fit without the 
LHC DY data included (shaded area) and the relative change 
in the ABM12 PDFs due to the LHC DY data obtained with one (solid line) 
and two (dashes) iterations of 
the fast algorithm used to take into account the DY NNLO corrections. 
The dotted lines display $1\sigma$ band for the ABM12 PDFs
obtained with one iteration of the algorithm. 
}
\end{figure}

\begin{figure}[th!]
\centerline{
  \includegraphics[width=16.0cm]{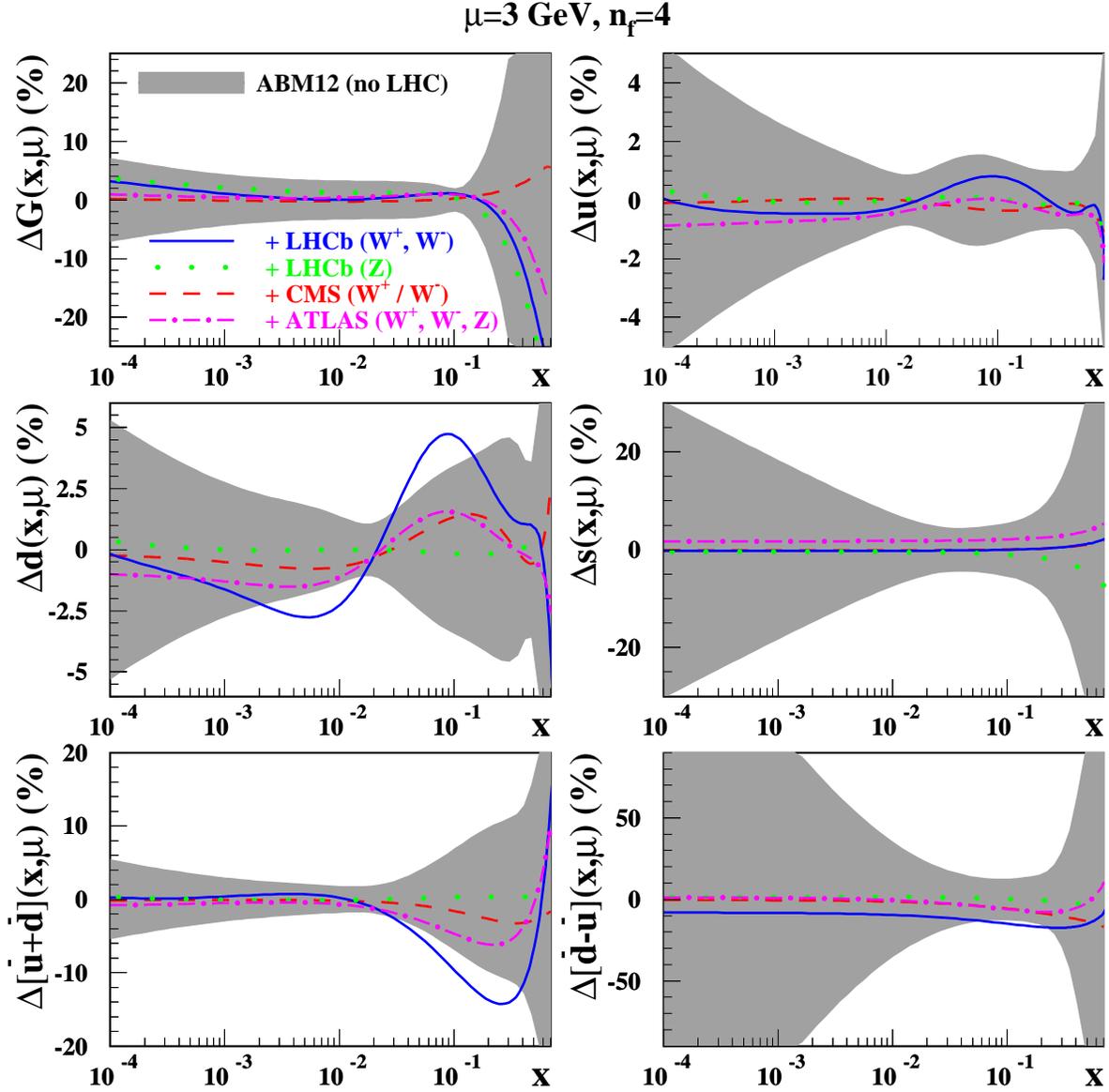}}
  \caption{\small
    \label{fig:lhc}
     The same as in Fig.~\ref{fig:abm12} for the variants of ABM12 fit
     including separate LHC DY data sets (sold line: LHCb~\cite{Aaij:2012vn},
     dots: LHCb~\cite{Aaij:2012mda},
     dashes: CMS~\cite{Chatrchyan:2012xt},
     dashed dots: ATLAS~\cite{Aad:2011dm}).
}
\end{figure}

\begin{figure}[t!]
\centerline{
  \includegraphics[width=16.0cm]{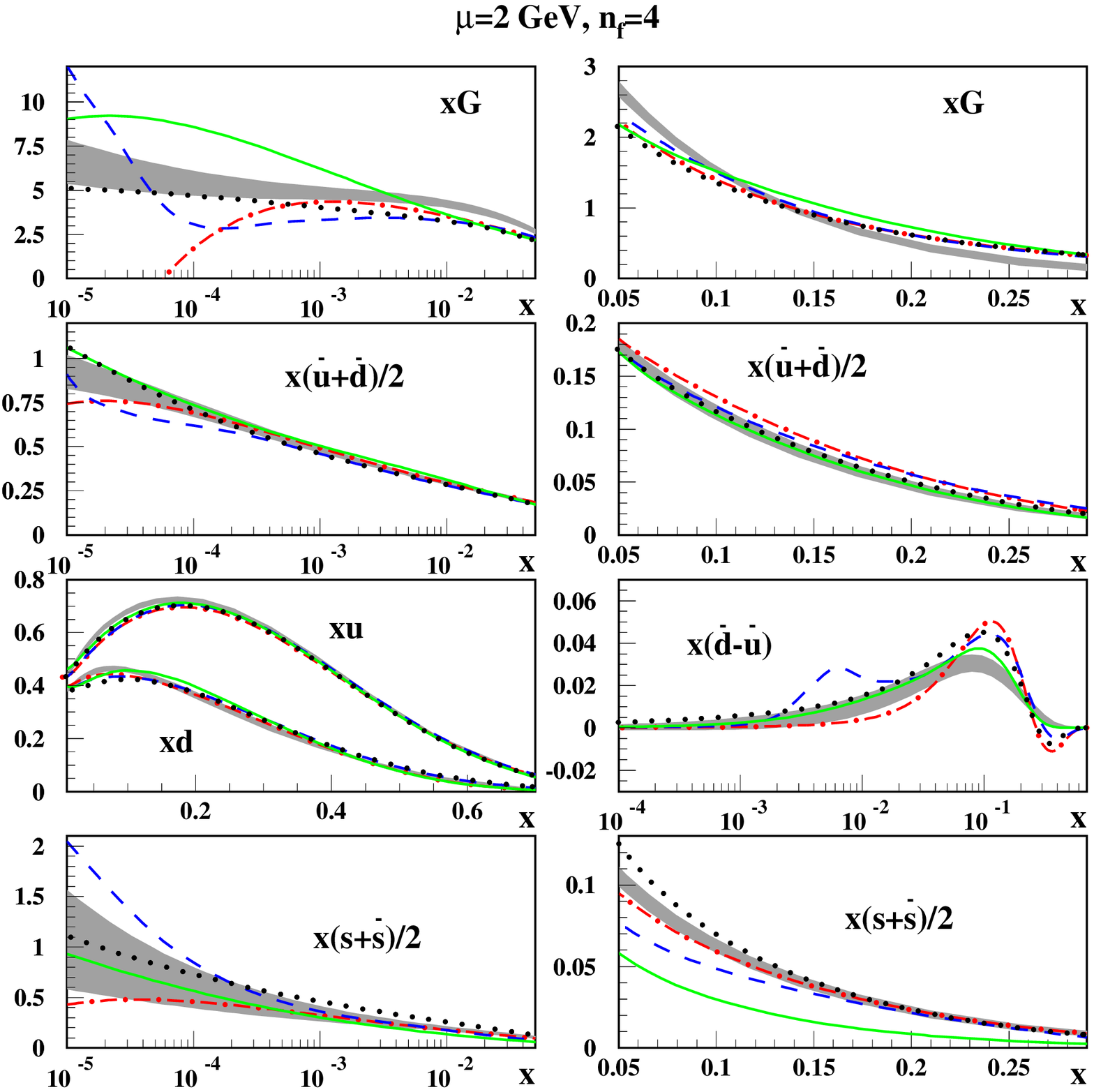}}
  \caption{\small
    \label{fig:pdfs}
    The 1$\sigma$ band for the 4-flavor NNLO ABM12 PDFs at the scale 
    of $\mu=2~{\rm GeV}$ versus $x$ obtained  in this analysis (shaded area) 
    compared with the ones obtained by other groups 
     (solid lines: JR09~\cite{JimenezDelgado:2008hf}, dashed dots: MSTW~\cite{Martin:2009iq}, 
    dashes: NN23~\cite{Ball:2012cx},
    dots: CT10~\cite{Gao:2013xoa}).
}
\end{figure}

\subsection{Comparison with ABM11 and other PDFs}
\label{sec:comp}

The PDFs obtained in the present analysis are basically in agreement with
the ABM11 ones obtained in the earlier version of our
fit~\cite{Alekhin:2012ig} within the uncertainties, cf. 
Fig.~\ref{fig:abm11}. 
The strange quark distribution is particularly stable 
since in our analysis it is constrained by the neutrino-induced dimuon
production that was not updated neither from the experimental nor from the 
theoretical side. It is still significantly suppressed as compared to the
non-strange sea and this contrasts with the strangeness enhancement   
found in the ATLAS PDF analysis
based on the collider data only~\cite{Aad:2011dm}.  
The change in the gluon distribution 
happens in particular due to impact of the HERA charm data and improvements in
the heavy-quark electro-production description, cf. Ref.~\cite{Alekhin:2012vu} 
for details. At the same time 
the ABM12 quark distributions differ from the ABM11 ones
at most due to the LHC DY data. 
This input contributes to a better separation of the non-strange 
sea and the valence quark distributions. 
As a result, at the factorization scale
$\mu=3~{\rm GeV}$ and $x\sim 0.2$ the non-strange 
sea goes down by somewhat 15\%, while 
the total $d$-quark distribution goes up by some 2\%, 
cf. Fig.~\ref{fig:abm12}. 
In turn, this improvement allows for a better accuracy of 
both, the sea and the valence distributions, in particular, of the $d$-quark one. 
This improvement is particularly valuable since 
the accuracy of the latter is limited in the case of DIS data due to the uncertainty in the nuclear
correction employed to describe the deuterium-target data. 
The LHCb data on $W^+$ and $W^-$ production~\cite{Aaij:2012vn} 
provide the biggest impact on the PDFs as compared to other LHC data, 
cf. Fig.~\ref{fig:lhc}, due to the forward kinematics probed in this
experiment. It is also worth noting that the gluon distribution is 
also sensitive to the existing LHC DY data and in the ABM fit 
they pull it somewhat up (down) at small (large) $x$.
However, in general, the changes are within the PDF uncertainties. 
This justifies our approach of using the set of 
PDF uncertainties to pre-calculate the NNLO DY cross-section grid 
and then to compute those cross sections by grid interpolation in minimal time. 
To provide the best accuracy of this algorithm 
the ABM12 PDFs are produced taking the DY cross-section grid 
calculated for the PDFs obtained in the variant of ABM12, which differs from 
the nominal ABM12 one by inclusion of the LHC data only.
Furthermore, to check explicitly the stability of the algorithm we perform 
a second iteration of the fit based on the DY cross-section grids prepared
with the PDFs obtained in the first iteration. 
The iterations demonstrate nice convergence and 
the first iteration suffices to obtain an accurate
result, cf. Fig.~\ref{fig:abm12}. 

The NNLO PDFs obtained in this analysis are compared to the results of 
other groups in Fig.~\ref{fig:pdfs}. Our PDFs are in reasonable agreement with
the newly released CT10 PDFs~\cite{Gao:2013xoa}. The most striking difference
is observed for the large-$x$ gluon distribution, which is constrained by the
Tevatron jet data in the CT10 analysis. It is worth noting that this 
constraint is obtained for
CT10 using the NLO corrections only, while the NNLO corrections 
may be as big as 15-25\%~\cite{Ridder:2013mf}. Therefore the discrepancy
between CT10 and our result should decrease once the NNLO corrections to the
jet production  are taken into
account. Comparison of the ABM12 PDFs with the ones obtained by other groups 
demonstrate the trend similar to 
the ABM11 case~\cite{Alekhin:2012ig}. The most essential difference 
appears in the large-$x$ gluon distribution. It is also constrained by the
Tevatron jet data for MSTW08~\cite{Martin:2009iq} and 
NN23~\cite{Ball:2012cx}, with the
NNLO corrections due to the threshold resummation taken into account 
in this case. However, the threshold resummation 
terms used in Refs.~\cite{Martin:2009iq,Ball:2012cx}
introduce additional theoretical uncertainties~\cite{Kumar:2013hia}.
Therefore, a conclusive comparison with our results is still impractical.
The spread in the small-$x$ 
gluon distribution obtained by different groups can be consolidated 
with the help of the 
H1 data on the structure function $F_L$~\cite{Adloff:2000qk} being sensitive in 
this region.
Similarly, differences in the estimates of the 
non-strange sea distribution at $x\sim 0.2$
can be eliminated using the LHC DY data considered in our analysis. 
At the same time the observed spread in the results for 
the strange sea shape cannot be explained by a particular data selection  
or difference in the theoretical accuracy of the analyses since 
all the groups use the CCFR and NuTeV data on the neutrino-induced 
dimuon production~\cite{Goncharov:2001qe} as a strange sea constraint
and take into account the NLO corrections to this 
process~\cite{Gluck:1996ve,Gottschalk:1980rv}. The very recent precise 
data on the neutrino-induced dimuon production by
NOMAD~\cite{Samoylov:2013xoa} are still not included in the present analysis.
However, they demonstrate good agreement with the ABM11 prediction and may
help to consolidate different estimates of the shape of the strange sea.

\subsection{The strong coupling constant and the charm quark mass}
\label{sec:const}

The strong coupling constant $\alpha_s(M_Z)$ is measured together with the 
parameters of the PDFs, the heavy-quark mass $m_c$ and the higher 
twist parameters within the analysis. The present accuracies of the scaling
violations of the deep-inelastic world data make the use of NNLO QCD corrections mandatory. 
At NLO the scale uncertainties typically amount to $O(5\%)$, cf.~\cite{Blumlein:1996gv},
and, therefore, are simply too large.

The value of $\alpha_s(M_Z)$ obtained in the present analysis is 
\begin{eqnarray}
\label{eq:as}
\alpha_s^{\rm NNLO}(M_Z) &=& 0.1132 \pm 0.0011 
\, .
\end{eqnarray}
This result is in excellent agreement with those given by other groups and
by us in 
Refs.~\cite{Alekhin:2009ni,Blumlein:2006be,Blumlein:2012se,Gluck:2006yz,JimenezDelgado:2008hf,Alekhin:2012ig},
see~Tab.~\ref{tab:TAB6}. 
As has been shown in \cite{Alekhin:2012ig} in detail the $\alpha_s$-values 
obtained upon analyzing the partial data sets from BCDMS \cite{Benvenuti:1989rh,Benvenuti:1989fm}, 
NMC \cite{Arneodo:1993kz,Arneodo:1996qe}, SLAC 
\cite{Whitlow:1990gk,Bodek:1979rx,Atwood:1976ys,Mestayer:1982ba,Gomez:1993ri,Dasu:1993vk},
HERA~\cite{Aaron:2009aa}, and the Drell-Yan data \cite{Towell:2001nh,Moreno:1990sf} 
both at NLO and NNLO do very well compare to each other and to the central value  within the experimental errors.

Fits including jet data have been carried out before both by JR \cite{Gluck:2007ck} and ABM 
\cite{Alekhin:2011cf,Alekhin:2012ig}, along with other groups, performing  
systematic studies including both jet data from the Tevatron and in \cite{Alekhin:2012ig} also 
from LHC~\footnote{Contrary statements given in Refs.~\cite{Ball:2012wy,Forte:2013wc} are incorrect; 
see Ref.~\cite{Alekhin:2012ig} for further details.}. 
We would like to note that it is very problematic to call 
present NNLO fits of the world DIS data including jet data NNLO analyses, since the 
corresponding jet scattering cross sections  are available at NLO {\it only}. 
The complete NNLO results for the corresponding jet cross sections have to be used in 
later analyses, since threshold resummations are not expected to deliver a sufficient 
description~\cite{Kumar:2013hia}~\footnote{Partial 
NNLO results on the hadronic di-jet cross section are available \cite{Ridder:2013mf}.}.

\begin{figure}[t!]
\centerline{
  \includegraphics[width=8.0cm]{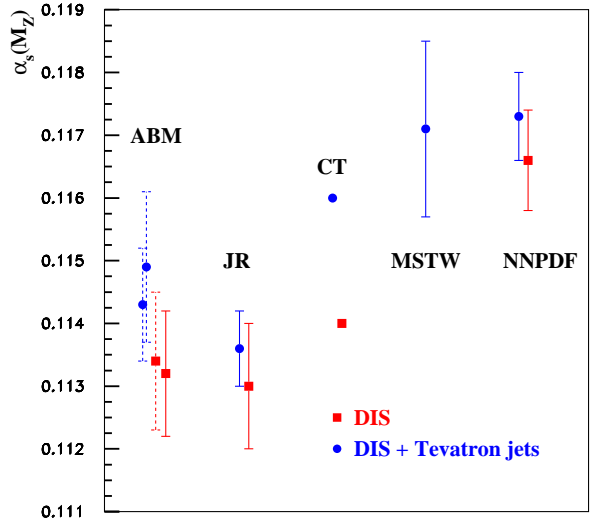}}
  \caption{\small
    \label{fig:alphas}
    The values of $\alpha_s(M_Z)$ at NNLO obtained in the PDF fits of ABM
    (solid bars: this analysis, dashed bars: ABM11~\cite{Alekhin:2012ig}) in 
    comparison with the  
    CT~\cite{Gao:2013xoa}, 
    JR~\cite{JimenezDelgado:2008hf}, 
    MSTW~\cite{Martin:2009iq} and 
    NNPDF~\cite{Ball:2011us} results. 
}
\end{figure}

The $\alpha_s(M_Z)$ values for some PDF groups are illustrated in Fig.~\ref{fig:alphas}. 
In Tab.~\ref{tab:TAB6} a general overview on the values of $\alpha_s(M_Z)$ at NNLO is given, with a few
determinations effectively at N$^3$LO in the valence analyses \cite{Blumlein:2006be,Blumlein:2012se},
and the hadronic $Z$-decay \cite{Baikov:2012er}. The BBG, BB, GRS, ABKM, JR, ABM11, CTEQ analyses and the
present analysis find lower values of $\alpha_s(M_Z)$ with errors at the 1--2\% level, while NN21 and MSTW08 
find larger values analyzing the deep-inelastic world data, Drell-Yan data, and partly also jet data
with comparable accuracy to the former ones.\footnote{Very recently MSTW \cite{THORNE:2013} reported lower 
values for $\alpha_s(M_z)$ also related to the LHC data.} 
Reasons for the higher values being obtained by NN21 and MSTW08 
were given in \cite{Alekhin:2012ig} before. As has been shown in \cite{Alekhin:2011ey} a consistent 
$F_L$-treatment for the NMC data and the BCDMS-data, cf.~\cite{Blumlein:2006be}, is necessary and leads to 
a change of the value of $\alpha_s(M_Z)$ to lower values. Furthermore, the sensitivity on kinematic cuts 
applied to remove higher twist effects has been studied. In the flavor non-singlet case this can be achieved 
by cutting for $W^2 > 12.5$~GeV$^2$, cf.~\cite{Blumlein:2006be}. In the singlet analysis there are also higher 
twist 
contributions in the lower $x$-region to be removed by applying the additional cut of $Q^2 > 10$ GeV$^2$, which, 
however, is not used by NN21 and MSTW08. We performed a fit without accounting for the higher twist terms and 
allowed for the range of data down to values of $Q^2 > 2.5$~GeV$^2$ at $W^2 > 12.5$~GeV$^2$,~\cite{Alekhin:2012ig}.
One obtains $\alpha_s(M_Z) = 0.1191 \pm 0.0016$, very close to the values found by NN21 and 
MSTW08. Comparisons of the $\alpha_s$ values in the fits by NN21 and MSTW08, furthermore, show strong 
variations with respect to different DIS data sets \cite{Alekhin:2012ig}, despite of the similar final value.

The analyses of thrust in $e^+e^-$ data by two groups also find low values, also with errors at the 1\% level. 
Higher values of $\alpha_s(M_Z)$ are obtained for the $e^+e^-$ 3-jet rate, the hadronic $Z$-decay, 
and $\tau$-decay within various analyses. The value of $\alpha_s(M_Z)$ has 
also been determined in different lattice simulations to high accuracy. The N$^3$LO values for 
$\alpha_s(M_Z)$ in the valence analyses \cite{Blumlein:2006be,Blumlein:2012se} yield slightly larger  
values than at NNLO. They are fully consistent with the NNLO values within errors. The corresponding
shift can be taken as a measure for the remaining theoretical uncertainty in the non-singlet-case, 
see~Tab.~\ref{tab:TAB6}.

Finally we would like to comment on recent determinations of $\alpha_s(M_Z)$ at NLO 
using the jet data~\cite{Malaescu:2012ts,Chatrchyan:2013txa}. 
The ATLAS and CMS jet data span a wider kinematic range than those of Tevatron 
and will allow very soon even more accurate measurements. In the analysis \cite{Chatrchyan:2013txa} 
$\alpha_s(M_Z)$ is 
determined scanning grids generated at different values of the strong coupling constants by the different
PDF-fitting groups. These are used to find a minimum for the jet data. Including the scale uncertainties the 
following NLO values are obtained for the 3/2 jet ratio by CMS \cite{Chatrchyan:2013txa}~:
\begin{eqnarray}
\label{jet1}
\alpha_s(M_Z) &=& 0.1148 \pm 0.0014~{(\rm exp.)} \pm 0.0018~{(\rm PDF)}
{\small \begin{array}{c} {+ 0.0050} \\ {- 0.0055} \end{array}}~{\rm (scale)} 
~~{\rm NNPDF21}~\mbox{\cite{Lionetti:2011pw}}
\, ,
 \\
\alpha_s(M_Z) &=& 0.1135 \pm 0.0018~{(\rm exp.)}~~[0.180~{(\rm favored~value)}]
~~{\rm CT10}~\mbox{\cite{Gao:2013xoa}} 
\, ,
\\
\alpha_s(M_Z) &=& 0.1141 \pm 0.0022~{(\rm exp.)}~~
[0.1202 {\small \begin{array}{c} +0.0012 \\ -0.0015 \end{array}}]
~~{\rm MSTW08}~\mbox{\cite{Martin:2009bu}}
\, . 
\label{jet3}
\end{eqnarray} 

A comparable NLO value has been reported using ATLAS jet data \cite{Malaescu:2012ts}
\begin{eqnarray} 
\alpha_s(M_Z) = 0.1151 \pm 0.0050~{\rm (exp.)} {\small \begin{array}{c} +0.0080 \\ 
-0.0073\end{array}}~{\rm (th.)}~.
\end{eqnarray} 

\begin{table}[h!]
\renewcommand{\arraystretch}{1.3}
\begin{center}
\begin{tabular}{|l|l|l|}
\hline
\multicolumn{1}{|c|}{ } &
\multicolumn{1}{c|}{$\alpha_s({M_Z})$} &
\multicolumn{1}{c|}{  } \\
\hline
BBG      & $0.1134~^{+~0.0019}_{-~0.0021}$
         & {\rm valence~analysis, NNLO}                  \cite{Blumlein:2006be}           
\\[0.5ex]
BB       & $0.1132  \pm 0.0022$
         & {\rm valence~analysis, NNLO}                  \cite{Blumlein:2012se}           
\\
GRS      & $0.112 $ & {\rm valence~analysis, NNLO}       \cite{Gluck:2006yz}           
\\
ABKM     & $0.1135 \pm 0.0014$ & {\rm HQ:~FFNS~$n_f=3$}  \cite{Alekhin:2009ni}             
\\
ABKM     & $0.1129 \pm 0.0014$ & {\rm HQ:~BSMN-approach} \cite{Alekhin:2009ni}             
\\
JR       & $0.1128 \pm 0.0010$ & {\rm dynamical~approach} \cite{JR:2013}   
\\
JR       & $0.1140 \pm 0.0006$ & {\rm including jet~data} \cite{JR:2013}
\\
ABM11            & $0.1134\pm 0.0011$ & \cite{Alekhin:2012ig}
\\
ABM12            & $0.1132\pm 0.0011$ & Eq.~(\ref{eq:as}) this work 
\\
MSTW & $0.1171\pm 0.0014$ &  \cite{Martin:2009bu}     
\\
MSTW & $0.1155-0.1175$ &  \cite{THORNE:2013}     
\\
NN21 & $0.1173\pm 0.0007$ &  \cite{Ball:2011us}     
\\
CTEQ & $0.1159 ... 0.1162$  &  \cite{Gao:2013xoa} 
\\
CTEQ & $0.1140$  &  (without jets) \cite{Gao:2013xoa} 
\\
BBG & $0.1141~^{+~0.0020}_{-~0.0022}$ & {\rm valence~analysis, N$^3$LO$(^*)$}  \cite{Blumlein:2006be}            
\\[0.5ex]
BB & $0.1137 \pm 0.0022$
& {\rm valence~analysis, N$^3$LO$(^*)$}  \cite{Blumlein:2012se}            
\\
\hline
{\rm $e^+e^-$~thrust} & $0.1140 \pm 0.0015$ & Abbate et al. \cite{Abbate:2012jh}
\\
{\rm $e^+e^-$~thrust} & $0.1131~^{+~0.0028}_{-~0.0022}$ & Gehrmann et al. \cite{Gehrmann:2012sc}
\\[0.5ex]
3 jet rate   & $0.1175 \pm 0.0025$ & Dissertori et al. 2009 \cite{Dissertori:2009qa}
\\
Z-decay      & $0.1189 \pm 0.0026$ & BCK 2008/12, N$^3$LO \cite{Baikov:2008jh,Baikov:2012er}
\\
$\tau$ decay & $0.1212 \pm 0.0019$ & BCK 2008               \cite{Baikov:2008jh}
\\
$\tau$ decay & $0.1204 \pm 0.0016$ & Pich 2011              \cite{Bethke:2011tr}
\\
$\tau$ decay & $0.1191 \pm 0.0022$ & Boito et al. 2012      \cite{Boito:2012cr}
\\
\hline
lattice      & $0.1205 \pm 0.0010$ & PACS-CS 2009 (2+1 fl.) \cite{Aoki:2009tf} 
\\
lattice      & $0.1184 \pm 0.0006$ & HPQCD 2010             \cite{McNeile:2010ji} 
\\
lattice      & $0.1200 \pm 0.0014$ & ETM 2012 (2+1+1 fl.)   \cite{Blossier:2012ef} 
\\
lattice      & $0.1156 \pm 0.0022$ & Brambilla et al. 2012 (2+1 fl.)  \cite{Bazavov:2012ka} 
\\
lattice      & $0.1181 \pm 0.0014$ & JLQCD   \cite{Shintani:2010ph} 
\\
\hline
{world average} & {$
0.1184 \pm 0.0007$  } & Bethke 2012 \cite{Bethke:2012jm} \\
\hline
\end{tabular}
\renewcommand{\arraystretch}{1}   
\caption{\small
Summary of recent NNLO and N$^3$LO QCD analyses of the DIS world data, supplemented by related measurements
using a series of other processes and lattice determinations. In case that jet data from hadron colliders are
used in the analysis the values of $\alpha_s(M_Z)$ cannot be considered NNLO values. 
\label{tab:TAB6}}
\end{center}
\end{table}

Interestingly, rather low values are obtained already at NLO. 
In parenthesis we quote the NLO values for $\alpha_s(M_Z)$ in Eqs.~(\ref{jet1})-(\ref{jet3}) 
which are obtained by the fitting groups at minimal $\chi^2$. Obviously the values found in the jet-data 
analysis do not correspond to these values. Yet, the NLO scale uncertainty in this analysis 
is large. Recently the jet energy-scale error has been improved by CMS \cite{Chatrchyan:2013txa}, 
leading to a significant reduction of the experimental error. The gluonic NNLO $K$-factor 
is positive; as shown in Fig.~2 of Ref.~\cite{Ridder:2013mf} the scale dependence for 
$\mu = \mu_F = \mu_R$ behaves flat over a wide range of scales. It is therefore expected that also the 
error due to scale variation will turn out to be very small in the NNLO analysis. 
It will be important to repeat this analysis and to fit the LHC jet data together with the world deep-inelastic data,
which will be also instrumental for the determination of the gluon distribution 
at large scales. 

The present DIS world data together with the $F_2^{c\bar{c}}(x,Q^2)$ data, are competitive in the 
determination of the charm quark mass in a correlated fit with the PDF-parameters and $\alpha_s(M_Z)$.
For the $\overline{\rm MS}$ mass the value of 
\begin{eqnarray}
\label{eq:mcv}
m_c(m_c) = 1.24 \pm 0.03~{\rm (exp.)} 
{\small 
\begin{array}{c} +0.04 \\ -0.00 \end{array}} {\rm (th.)}~{\rm GeV}
\end{eqnarray}
is obtained at NNLO, see also \cite{Alekhin:2012vu}. At present this analysis is the only one, in 
which all known higher order heavy-flavor corrections to deep-inelastic scattering have been considered. 
This value still should be quoted as of approximate NNLO, since the NNLO-corrections are only modeled
\cite{Kawamura:2012cr} combining small-$x$ and threshold resummation effects with information of the 
3-loop moments of the heavy-flavor Wilson coefficients \cite{Bierenbaum:2009mv} at high values of $Q^2$.
Two scenarios have been considered in \cite{Kawamura:2012cr} to parameterize the Wilson coefficients
accounting for an estimated error. Here the fit favors a region of the parameter $d_N \in [-0.1, 0.5]$, 
cf. \cite{Kawamura:2012cr}, on which the theoretical error is based~\footnote{The calculation of the 
exact NNLO heavy-flavor Wilson coefficients is underway 
\cite{Ablinger:2010ty,Blumlein:2012vq,Ablinger:2012qm,Blumlein:2013ota}.}. 
The value in Eq.~(\ref{eq:mcv}) compares well to the present world average of 
$m_c(m_c) = 1.275 \pm 0.025$~GeV~\cite{Beringer:1900zz}.

It is needless to say that the determination of a fundamental parameter of the SM, such as 
$m_c$, has to follow a thorough quantum field-theoretic prescription, rather than specific models
also being found in the literature, cf., e.g.~\cite{Thorne:2008xf}. Despite the correct renormalization 
procedure of the heavy-flavor Wilson coefficients to 3-loop order were known \cite{Bierenbaum:2009mv} to 
3-loop order, there are still even massless scenarios to this level, cf. e.g. \cite{Stavreva:2012bs}, which 
ignore the exact theoretical description.

\subsection{Standard candle cross sections}
\label{sec:standard-candles}
In this Section we quantify the impact of the new PDF set on the predictions for
benchmark cross sections at the LHC for various c.m.s. energies.
To that end, we confine ourselves to (mostly) inclusive cross sections which are known
to NNLO in QCD, see~\cite{Alekhin:2012ig,Alekhin:2010dd} for previous benchmark numbers, 
since the NNLO accuracy is actually the first instance, where meaningful statements 
about the residual theoretical uncertainty are possible given the precision of
present collider data and the generally large residual variation of the renormalization 
and factorization scale at NLO.

In detail, we consider the following set of inclusive observables at NNLO in QCD:
hadronic $W$- and $Z$-boson production~\cite{Hamberg:1990np,Harlander:2002wh}, the cross section
for Higgs boson production in gluon-gluon fusion~\cite{Harlander:2002wh,Anastasiou:2002yz,Ravindran:2003um,Ravindran:2004mb},
and the cross section for top-quark pair 
production~\cite{Baernreuther:2012ws,Czakon:2012zr,Czakon:2012pz,Czakon:2013goa,Langenfeld:2009wd}.
We have used the {\tt LHAPDF} library~\cite{Whalley:2005nh,lhapdf:2013} 
for the cross section computations to interface to our PDFs provided in the form of data grids 
for $n_f=3,4$ and $5$ flavors accessible with the {\tt LHAPDF} library~\footnote{
The {\tt LHAPDF} library can be obtained from {\tt http://projects.hepforge.org/lhapdf} together with installation instructions.}, 
\begin{verbatim}
      abm12lhc_3_nnlo.LHgrid (0+28),
      abm12lhc_4_nnlo.LHgrid (0+28),
      abm12lhc_5_nnlo.LHgrid (0+28),
\end{verbatim}
which contains the central fit and 28 additional sets for the combined symmetric
uncertainty on the PDFs, on $\alpha_s$ and on the heavy-quark masses.
All PDF uncertainties quoted here are calculated in the standard manners, i.e., 
as the $\pm 1\sigma$-variation.

\subsubsection{$W$- and $Z$-boson production}
\label{sec:wzprod}

We start by presenting results for $W$- and $Z$-boson production at the LHC.
For the electroweak parameters, we follow~\cite{Alekhin:2012ig,Alekhin:2010dd} 
and choose the scheme based on the set $(G_F, M_W, M_Z)$.
According to~\cite{Beringer:1900zz}, we have 
$G_F = 1.16638 \times 10^{-5}~{\rm GeV}^{-2}$, 
$M_W = 80.385  \pm 0.015$~GeV, 
$M_Z = 91.1876  \pm 0.0021$~GeV 
and the corresponding widths 
$\Gamma(W^\pm) = 2.085 \pm 0.042$~GeV and 
$\Gamma(Z) = 2.4952 \pm 0.0023$~GeV.
The weak mixing angle is then a dependent quantity, with  
\begin{equation}
  \label{eq:sintw}
  \hat{s}_Z^2 \,=\,  1 - \frac{M_W^2}{\hat{\rho} M_Z^2}  = 0.23098 \pm 0.00041
  \, ,
\end{equation}
and $\hat{\rho} = 1.01051 \pm 0.00011$.
The Cabibbo angle $\theta_c$ yields the value of $\sin^2 \theta_c = 0.05085$.

The change in the predictions between ABM11 and ABM12 is small and for the 
current theoretical accuracy, the uncertainty due to the scale variation is
already significantly smaller compared to the PDF error, see Tabs.~\ref{tab:wz7}--\ref{tab:wz14}.
This indicates the very good stability of the PDF fit and the consistency of the 
previous ABM11 PDFs with the new variant including LHC data.
An additional source of theoretical uncertainty for $W$- and $Z$-boson production, 
namely the choice of PDF sets with $n_f=4$ or with $n_f=5$ flavors (as in Tabs.~\ref{tab:wz7}--\ref{tab:wz14})
has already been discussed and quantified in~\cite{Alekhin:2012ig}.
Generally, those differences are less than 1$\sigma$ in the PDF uncertainty
and become successively smaller as perturbative corrections of higher order
are included.

\begin{table}[ht!]
\renewcommand{\arraystretch}{1.3}
\begin{center}
{\small
\begin{tabular}{|l|l|l|l|l|}
\hline
LHC7
&$W^+$ & $W^-$ & $W^\pm$ & $Z$ \\[0.5ex]
\hline 
ABM11 &
${59.53}~^{+0.38}_{-0.23}~^{+0.88}_{-0.88}$
 &
${39.97}~^{+0.28}_{-0.17}~^{+0.65}_{-0.65}$
 &
${99.51}~^{+0.69}_{-0.41}~^{+1.43}_{-1.43}$
 &
${29.23}~^{+0.18}_{-0.10}~^{+0.42}_{-0.42}$ 
 \\[0.5ex]
ABM12 &
${58.40}~^{+0.38}_{-0.24}~^{+0.70}_{-0.70}$
 &
${39.63}~^{+0.29}_{-0.18}~^{+0.45}_{-0.45}$
 &
${98.03}~^{+0.67}_{-0.41}~^{+1.13}_{-1.13}$
 &
${28.79}~^{+0.17}_{-0.11}~^{+0.33}_{-0.33}$
 \\[0.5ex]
\hline
\end{tabular}
}
\caption{\small 
  \label{tab:wz7}
  The total cross sections [pb] for gauge boson production 
  at the LHC with $\sqrt{s} = 7~{\rm TeV}$ 
  for the $n_f=5$ flavor PDF sets ABM11 and ABM12 at NNLO accuracy.
  The errors shown are the scale uncertainty based 
  on the shifts $\mu=M_{W/Z}/2$ and $\mu = 2M_{W/Z}$ 
  and, respectively, the 1$\sigma$ PDF uncertainty.  
}
\end{center}
\end{table}
\begin{table}[ht!]
\renewcommand{\arraystretch}{1.3}
\begin{center}
{\small
\begin{tabular}{|l|l|l|l|l|}
\hline
LHC8
&$W^+$ & $W^-$ & $W^\pm$ & $Z$ \\[0.5ex]
\hline
ABM11 &
${68.30}~^{+0.48}_{-0.29}~^{+1.02}_{-1.02}$
 &
${46.67}~^{+0.35}_{-0.22}~^{+0.748}_{-0.748}$
 &
${114.97}~^{+0.82}_{-0.51}~^{+1.67}_{-1.67}$
 &
${33.97}~^{+0.23}_{-0.13}~^{+0.50}_{-0.50}$
 \\[0.5ex]
ABM12 &
${67.03}~^{+0.48}_{-0.30}~^{+0.81}_{-0.81}$
 &
${46.27}~^{+0.35}_{-0.23}~^{+0.53}_{-0.53}$
 &
${113.29}~^{+0.84}_{-0.52}~^{+1.32}_{-1.32}$
 &
${33.49}~^{+0.22}_{-0.14}~^{+0.38}_{-0.38}$
 \\[0.5ex]
\hline
\end{tabular}
}
\caption{\small 
  \label{tab:wz8}
The same as Tab.~\ref{tab:wz7} for the LHC with $\sqrt{s} = 8~{\rm TeV}$.
}
\end{center}
\end{table}
\begin{table}[ht!]
\renewcommand{\arraystretch}{1.3}
\begin{center}
{\small
\begin{tabular}{|l|l|l|l|l|}
\hline
LHC13
&$W^+$ & $W^-$ & $W^\pm$ & $Z$ \\[0.5ex]
\hline
ABM11 &
${110.77}~^{+0.97}_{-0.61}~^{+1.70}_{-1.70}$
 &
${80.02}~^{+0.72}_{-0.47}~^{+1.28}_{-1.28}$
 &
${190.79}~^{+1.68}_{-1.09}~^{+2.87}_{-2.87}$
 &
${57.62}~^{+0.48}_{-0.29}~^{+0.88}_{-0.88}$
 \\[0.5ex]
ABM12 &
${108.86}~^{+0.97}_{-0.62}~^{+1.39}_{-1.39}$
 &
${79.33}~^{+0.73}_{-0.48}~^{+0.95}_{-0.95}$
 &
${188.19}~^{+1.69}_{-1.09}~^{+2.31}_{-2.31}$
 &
${56.88}~^{+0.48}_{-0.29}~^{+0.69}_{-0.69}$
 \\[0.5ex]
\hline
\end{tabular}
}
\caption{\small 
  \label{tab:wz13}
The same as Tab.~\ref{tab:wz7} for the LHC with $\sqrt{s} = 13~{\rm TeV}$.
}
\end{center}
\end{table}
\begin{table}[ht!]
\renewcommand{\arraystretch}{1.3}
\begin{center}
{\small
\begin{tabular}{|l|l|l|l|l|}
\hline
LHC14
&$W^+$ & $W^-$ & $W^\pm$ & $Z$ \\[0.5ex]
\hline
ABM11 &
${119.03}~^{+1.07}_{-0.68}~^{+1.83}_{-1.83}$
 &
${86.63}~^{+0.80}_{-0.53}~^{+1.39}_{-1.39}$
 &
${205.66}~^{+1.87}_{-1.20}~^{+3.12}_{-3.12}$
 &
${62.31}~^{+0.53}_{-0.32}~^{+0.96}_{-0.96}$
 \\[0.5ex]
ABM12 &
${116.99}~^{+1.07}_{-0.69}~^{+1.51}_{-1.51}$
 &
${85.89}~^{+0.80}_{-0.53}~^{+1.04}_{-1.04}$
 &
${202.88}~^{+1.87}_{-1.22}~^{+2.52}_{-2.52}$
 &
${61.52}~^{+0.53}_{-0.33}~^{+0.75}_{-0.75}$
 \\[0.5ex]
\hline
\end{tabular}
}
\caption{\small 
  \label{tab:wz14}
The same as Tab.~\ref{tab:wz7} for the LHC with $\sqrt{s} = 14~{\rm TeV}$.
}
\end{center}
\end{table}

\subsubsection{Higgs boson production}
\label{sec:higgsprod}

Let us now discuss the cross section for the SM Higgs boson
production in the gluon-gluon fusion channel, which is predominantly 
driven by the gluon PDF and the value of $\alpha_s(M_Z)$ from the effective vertex.
The known NNLO QCD corrections~\cite{Harlander:2002wh,Anastasiou:2002yz,Ravindran:2003um,Ravindran:2004mb} 
still lead to a sizable increase in 
\begin{table}[th!]
\renewcommand{\arraystretch}{1.3}
\begin{center}
{\small
\hspace*{-5mm}
\begin{tabular}{|l|l|l|l|l|l|}
\hline
     &
LHC7 &
LHC8 &
LHC13 &
LHC14 
 \\[0.5ex]
\hline
ABM11 &
$ 13.23~_{- 1.31}^{+ 1.35}~^{+ 0.30}_{- 0.30}$
 &
$ 16.99~_{- 1.63}^{+ 1.69}~^{+ 0.37}_{- 0.37}$
 &
$ 39.57~_{-3.42}^{+3.60}~^{+0.77}_{-0.77}$
 &
$ 44.68~_{- 3.78}^{+ 4.02}~^{+ 0.85}_{- 0.85}$ 
 \\[0.5ex]
ABM12 &
$ 13.28~_{-1.32}^{+1.35}~^{+0.31}_{-0.31}$
 &
$ 17.05~_{-1.64}^{+1.68}~^{+0.39}_{-0.39}$
 &
$ 39.69~_{-3.42}^{+3.60}~^{+0.84}_{-0.84}$
 &
$ 44.81~_{-3.80}^{+4.01}~^{+0.94}_{-0.94}$
 \\[0.5ex]
\hline
\end{tabular}
}
\caption{\small 
The total Higgs production cross sections [pb] in gluon-gluon fusion 
for the PDF sets ABM11 and ABM12 at NNLO accuracy using a Higgs boson mass $m_H = 125~{\rm GeV}$.
The errors shown are the scale uncertainty based 
on the shifts $\mu=m_H/2$ and $\mu = 2m_H$ and, respectively, the 1$\sigma$ PDF uncertainty.  
}
\label{tab:higgs}
\end{center}
\end{table}
\begin{table}[th!]
\renewcommand{\arraystretch}{1.3}
\begin{center}
{\small
\hspace*{-5mm}
\begin{tabular}{|l|l|l|l|l|l|}
\hline
     &
LHC7 &
LHC8 &
LHC13 &
LHC14 
 \\[0.5ex]
\hline
HiggsXSWG~\cite{Heinemeyer:2013tqa} &
$ 15.13~_{- 1.18}^{+ 1.07}~^{+ 1.15}_{- 1.07}$
 &
$ 19.27~_{- 1.50}^{+ 1.39}~^{+ 1.45}_{- 1.33}$
 &
$-$
 &
$ 49.85~_{- 4.19}^{+ 6.08}~^{+ 3.69}_{- 3.09}$ 
 \\[0.5ex]
\hline
\end{tabular}
}
\caption{\small 
The total Higgs production cross sections [pb] in gluon-gluon fusion 
of~\cite{Heinemeyer:2013tqa} used by ATLAS and CMS for a Higgs boson mass $m_H = 125~{\rm GeV}$.
The errors shown are the scale uncertainty and, respectively, the PDF+$\alpha_s$ uncertainty.  
}
\label{tab:higgsXSWG}
\end{center}
\end{table}

\noindent
the cross section at nominal values of the scale, i.e. $\mu=m_H$, 
and it is well established that a further stabilization beyond NNLO may be achieved 
on the basis of soft gluon resummation, see e.g.,~\cite{Moch:2005ky}.
At NNLO accuracy in QCD the theoretical uncertainty from the scale variation is
dominating by far over the PDF uncertainty.
Using a Higgs boson mass $m_H = 125~{\rm GeV}$ in Tab.~\ref{tab:higgs} 
we observe again only rather small changes between the ABM11 and the ABM12 predictions. 
This demonstrates that the gluon PDF is well constrained from existing data
and that the ABM11 results are consistent with the new fit based on including
selected LHC ones.

It is therefore interesting to compare the ABM predictions in
Tab.~\ref{tab:higgs} to the cross section values recommended for use in the
ongoing ATLAS and CMS Higgs analyses~\cite{Heinemeyer:2013tqa},
cf., Tab.~\ref{tab:higgsXSWG}~\footnote{See also {\tt
    https://twiki.cern.ch/twiki/bin/view/LHCPhysics/CERNYellowReportPageAt7TeV}
  for details.}.
The central values of the ABM predictions are significantly lower by some 11-14 \%. 
Only a small fraction of this difference can be attributed to the inclusion of 
soft gluon resummation beyond NNLO, which typically does reduce the scale
dependence, though, as is obvious from Tab.~\ref{tab:higgsXSWG}, 
and to the inclusion of other quantum corrections in~\cite{Heinemeyer:2013tqa}, e.g., the electro-weak ones.
Much larger sensitivity of the Higgs cross section predictions arise from
theory assumptions made in the analyses, e.g., 
for constraints from higher orders in QCD due to the treatment of fixed-target DIS data, see~\cite{Alekhin:2011ey}.
The most interesting aspect is the fact, that the PDFs+$\alpha_s$ error 
in~\cite{Heinemeyer:2013tqa} is inflated roughly by a factor of 4 in comparison 
to our predictions in Tab.~\ref{tab:higgs}, where we quote the 
1$\sigma$ PDF (and $\alpha_s$ of course) error entirely determined 
from the correlated experimental uncertainties in the fitted data.
In summary, the cross section predictions~\cite{Heinemeyer:2013tqa} used in 
the current Higgs analyses at the LHC are subject to both, a bias due 
to specific theory assumptions made in PDF and $\alpha_s$ fits as well as 
largely overestimated uncertainties of the relevant non-perturbative input.
Thus, checks of correlations between experimental data for different scattering processes 
at the LHC and their sensitivity to PDFs along the lines of
Sec.~\ref{sec:basics} are urgently needed to consolidate this issue, cf.~\cite{Dittmaier:2012vm}.

\subsubsection{Top-quark pair production}
\label{sec:ttbar-nums}

Finally, we present predictions for the total cross section for 
$t{\bar t}$-pair hadro-production in Tabs.~\ref{tab:ttbar-pole} and \ref{tab:ttbar-msbar}.
Using the program {\tt Hathor} (version 1.5)~\cite{Aliev:2010zk} which
incorporates the recently completed QCD corrections at NNLO~\cite{Baernreuther:2012ws,Czakon:2012zr,Czakon:2012pz,Czakon:2013goa}, 
we give numbers for two representative top-quark masses, that is the 
running mass $m_t(m_t)=162$~GeV in the $\overline{\rm MS}$ mass scheme 
and the pole mass $m_t({\rm pole})=171$~GeV in the on-shell scheme.

\begin{table}[th!]
\renewcommand{\arraystretch}{1.3}
\begin{center}
{\small
\hspace*{-5mm}
\begin{tabular}{|l|l|l|l|l|l|}
\hline
&\multicolumn{1}{|l|}{LHC7 }
&\multicolumn{1}{|l|}{LHC8 }
&\multicolumn{1}{|l|}{LHC13 }
&\multicolumn{1}{|l|}{LHC14 }
\\[0.5ex]     
\hline
ABM11 &
$141.6~^{+5.5}_{-8.7}~^{+6.9}_{-6.9}$ &
$207.2~^{+7.8}_{-12.5}~^{+9.3}_{-9.3}$ &
$727.6~^{+24.4}_{-39.4}~^{+23.7}_{-23.7}$ &
$867.8~^{+28.5}_{-46.3}~^{+27.0}_{-27.0}$
\\[0.5ex]
ABM12 &
$143.0~^{+5.6}_{-8.8}~^{+6.5}_{-6.5}$ &
$209.1~^{+7.9}_{-12.6}~^{+8.7}_{-8.7}$ & 
$732.2~^{+24.4}_{-39.6}~^{+22.9}_{-22.9}$ &
$872.9~^{+28.6}_{-46.3}~^{+26.2}_{-26.2}$
\\[0.5ex]
\hline
\end{tabular}
}
\caption{\small 
The total cross section for top-quark pair-production at NNLO [pb] using a 
pole mass $m_t({\rm pole}) = 171~{\rm GeV}$ and the PDF sets 
ABM11 and ABM12 and with the errors shown as $\sigma + \Delta \sigma_{\rm scale} + \Delta \sigma_{\rm PDF}$.
The scale uncertainty $ \Delta \sigma_{\rm scale}$ is based on maximal and minimal 
shifts for the choices $\mu=m_t({\rm pole})/2$ and $\mu = 2m_t({\rm pole})$ and 
$\Delta \sigma_{\rm PDF}$ is the 1$\sigma$ combined PDF+$\alpha_s$ error.
}
\label{tab:ttbar-pole}
\end{center}
\end{table}

\begin{table}[th!]
\renewcommand{\arraystretch}{1.3}
\begin{center}
{\small
\hspace*{-5mm}
\begin{tabular}{|l|l|l|l|l|l|}
\hline
&\multicolumn{1}{|l|}{LHC7 }
&\multicolumn{1}{|l|}{LHC8 }
&\multicolumn{1}{|l|}{LHC13 }
&\multicolumn{1}{|l|}{LHC14 }
\\[0.5ex]     
\hline
ABM11 &
$148.6~^{+0.2}_{-4.5}~^{+6.6}_{-6.6}$ &
$217.2~^{+0.2}_{-6.5}~^{+8.8}_{-8.8}$ &
$760.0~^{+0.0}_{-21.0}~^{+22.2}_{-22.2}$ &
$906.0~^{+0.0}_{-24.7}~^{+25.2}_{-25.2}$
\\[0.5ex]
ABM12 &
$150.2~^{+0.1}_{-4.6}~^{+6.1}_{-6.1}$ & 
$219.3~^{+0.1}_{-6.6}~^{+8.2}_{-8.2}$ & 
$765.1~^{+0.0}_{-21.1}~^{+21.3}_{-21.3}$ &
$911.6~^{+0.0}_{-24.7}~^{+24.4}_{-24.4}$
\\[0.5ex]
\hline
\end{tabular}
}
\caption{\small 
The same as Tab.~\ref{tab:ttbar-pole} for a running mass 
$m_t(m_t) = 162~{\rm GeV}$ in the \msbar\ scheme.
}
\label{tab:ttbar-msbar}
\end{center}
\end{table}

At NNLO accuracy in QCD, the PDF uncertainties given in
Tabs.~\ref{tab:ttbar-pole} and \ref{tab:ttbar-msbar} are dominating 
in comparison to the theory uncertainties based on scale variation.
As discussed at length in Sec.~\ref{sec:ttbar} the LHC data 
for $t{\bar t}$-pair production included in the ABM12 fit predominantly
constrains the top-quark mass and has little impact on the gluon PDF and on
the value of the strong coupling constant $\alpha_s(M_Z)$.
Therefore the cross section predictions of the ABM11 and ABM12 PDFs largely coincide.

\section{Conclusions}
\label{sec:conclusions}

We have presented the PDF set ABM12, which results from a global analysis of
DIS and hadron collider data including, for the first time, the available LHC data 
for the standard candle processes such as $W^\pm$- and $Z$-boson and $t{\bar t}$-production.
The analysis has been performed at NNLO in QCD and along with the new data included 
also progress in theoretical predictions has been reflected accordingly.
The new ABM12 analysis demonstrates very good consistency with the 
previous PDF sets (ABM11, ABKM09) regarding the parameter values for PDFs 
as well as the strong coupling constant $\alpha_s(M_Z)$ and the quark masses.
Continuous checks for the compatibility of the data sets along with a detailed account 
of the systematic errors and of the correlations among the fit parameters have
been of paramount importance in this respect.

In detail, we have considered new HERA data sets on semi-inclusive charm production in DIS 
in Sec.~\ref{sec:herac} which have allowed to validate the $c$-quark
production mechanism in the FFN scheme relying on 3 light flavors in the initial state 
and leading to a precise determination of the running $c$-quark mass.
As another new DIS data set, the neutral-current inclusive data at high $Q^2$
from HERA has been included, which exhibits sensitivity to the exchange of photons, $Z$-bosons as well as 
to $\gamma$-$Z$-interference. 
Our analysis in Sec.~\ref{sec:heraq2} corroborates again the fact, 
that even at high scales the FFN scheme is sufficient for description of the
DIS data.

The fit of LHC precision data on $W^\pm$- and $Z$-boson production 
improves the determination on the quark distributions at $x \sim 0.1$ and 
constrains especially the $d$-quark distribution.
The fit shows good consistency and a further reduction of the experimental systematic uncertainties 
would certainly strengthen the impact of the LHC DY data in global fits.
On the technical side, we remark that the fit of DY data has been based on the exact NNLO
differential cross section predictions, expanded over the set of eigenfunctions 
spanning the basis for the ABM PDF uncertainties.
This has served as a starting point for a rapidly converging fit including the LHC DY
data with account of all correlations.

Also data for the total $t{\bar t}$-cross section has been smoothly accommodated into the fit. 
A proper treatment of the correlation between the gluon PDF, the strong
coupling constant $\alpha_s(M_Z)$ and the top-quark mass has been crucial here.
Moreover, the running-mass definition for the top-quark provides a better description 
of data as compared to the pole mass case, the latter showing still sizable 
sensitivity to perturbative QCD corrections beyond NNLO accuracy.
Our analysis in Sec.~\ref{sec:ttbar} yields a precise value 
with an uncertainty of roughly 1.5 \% for the \msbar\ mass $m_t(m_t)$ which 
has been used to extract $m_t({\rm pole})$ at NNLO.

In summary, the new ABM12 fit demonstrates, that a smooth extension 
of the ABM global PDF analysis to incorporate LHC data is feasible 
and does not lead to large changes in the fit results.
As we have shown in Sec.~\ref{sec:comp} differences with respect to other PDFs sets remain.
However, these differences are based either on a different treatment of the data sets or on 
different theoretical descriptions of the underlying physical processes and we
have commented on the correctness of some of those procedures.
In particular, the value of strong coupling constant $\alpha_s(M_Z)$ 
in our analysis remains largely unchanged as documented in Sec.~\ref{sec:const}
and the theoretical predictions for benchmark cross section at the LHC are very stable.
This particularly applies to the cross section for Higgs production in
the gluon-gluon fusion shown in Sec.~\ref{sec:standard-candles}. We 
commented on the implications for the ongoing Higgs analyses at the LHC.

The precision of the currently available experimental data make global analyses at NNLO accuracy in QCD mandatory.
This offers the great opportunity for high precision determinations of the non-perturbative parameters 
relevant in theory predictions of hadron collider cross section.
At the same time, the great sensitivity to the underlying theory 
allows to test and to scrutinize remaining model prescriptions and, eventually, to reject wrong 
assumptions.

\subsection*{Acknowledgments}
We would like to thank H.~B\"ottcher for discussions, P.~Jimenez-Delgado  and E.~Reya 
for a private communication prior to publication.
We gratefully acknowledge the continuous support of M.~Whalley to integrate 
the results of the new ABM fit into the {\tt LHAPDF} library~\cite{Whalley:2005nh,lhapdf:2013}. 

J.B. acknowledges support from Technische Universit\"at Dortmund.
This work has been supported by Helmholtz Gemeinschaft under contract VH-HA-101 ({\it Alliance Physics at the Terascale}),
by Deutsche Forschungsgemeinschaft in Sonderforschungs\-be\-reich/Transregio~9, 
by Bundesministerium f\"ur Bildung und Forschung through contract (05H12GU8),
and by the European Commission through contract PITN-GA-2010-264564 ({\it LHCPhenoNet}).

\smallskip

\noindent
{\bf Note added:}
While this work was being finalized, a new combination of measurements of the top-quark pair production cross section from the
Tevatron appeared~\cite{Aaltonen:2013wca}, which carries a combined experimental uncertainty of 5.4\%.
This measurements yields $\sigma_{pp \to t{\bar t}} = 7.82 \pm 0.42$~pb
for a value of $m_t({\rm pole}) = 171$~GeV for the top-quark pole mass,
which is consistent with the NNLO cross section prediction 
$\sigma_{pp \to t{\bar t}} = 7.17~^{+0.22}_{-0.31}~^{+0.16}_{-0.16}$~pb based 
on the ABM12 PDFs at NNLO within the uncertainties.

\appendix
\renewcommand{\theequation}{\ref{sec:appA}.\arabic{equation}}
\setcounter{equation}{0}
\renewcommand{\thefigure}{\ref{sec:appA}.\arabic{figure}}
\setcounter{figure}{0}
\renewcommand{\thetable}{\ref{sec:appA}.\arabic{table}}
\setcounter{table}{0}
\section{A fast algorithm for involved computations in PDF fits}
\label{sec:appA}

The accommodation of the different data sets for the PDF fit demands very involved 
computations of the QCD corrections to the Wilson coefficients. In particular this 
applies to the calculation of the rapidity distribution of the $W$- and $Z$-boson decay 
products produced in hadronic collisions, which are based on the fully exclusive NNLO 
codes {\tt DYNNLO}\cite{Catani:2009sm} and {\tt FEWZ}~\cite{Li:2012wna}. The typical 
CPU run-time needed to achieve a 
calculation accuracy of much better than the uncertainty of the present data using the 
codes~\cite{Catani:2009sm,Li:2012wna} amounts to $O(100)$ hours. Therefore an iterative 
use of the available fully exclusive DY codes in the QCD fit is widely impossible.
Instead, these codes are commonly run in advance for the variety of PDF sets,
covering the foreseeable spread in the PDF variation, the results of which are stored 
grids. Afterwards the cross section values for a given PDF set can be computed in a fast manner
using linear grid interpolations. For the first time this approach was 
implemented in the code {\tt fastNLO} \cite{Wobisch:2011ij} for the NLO corrections to the jet 
productions cross sections. A similar approach is also used in the code 
{\tt AppleGrid}~\cite{Carli:2010rw} which provides a tool for generating the cross section grids of 
different processes, including the DY process. Since {\tt fastNLO} and {\tt AppleGrid} are 
tools of general purpose, the PDF basis used to generate those grids need to be sufficiently 
wide to cover the differences between the existing PDF sets. Meanwhile the possible variations of 
the PDFs in a particular fit are not very large, i.e. if a new fit is aimed to accommodate a new 
data set being in sufficient agreement with those used in earlier versions of the fit, 
one may expect variations of the PDFs being comparable to their uncertainties. In this case the 
PDF basis used to generate the grids for the cross section can be reliably selected as a 
PDF-bunch, 
which encodes the uncertainties in a given PDF set. For the PDF uncertainties estimated with the 
Hessian method this bunch is provided by the PDF set members corresponding to the $1 \sigma$
variation in the fitted parameters. This allows to minimize the size of the pre-calculated cross 
section grids and reduces the CPU time necessary to generate these grids correspondingly. 
Moreover, the structure of the calculation algorithm in using these grids for the PDF fit turns
out to be simple. In this appendix we describe, how this approach is implemented in the present 
analysis. 

Firstly, we remind the basics of the PDF uncertainty handling, see Ref.~\cite{Alekhin:2012ig} for 
details. Let $\vec q(P_i)$ be the vector of parton distributions encoding the gluon 
and 
quark species. It depends on the PDF parameters ${P_i}$ with the index $i \in [1, N_p]$ and $N_p$ 
the number of parameters. ${P_i^0}$ denote the parameter values obtained in the PDF fit and 
$\Delta{P_i}$ are their standard deviations. In general the errors in the parameters are 
correlated, which is expressed by a non-diagonal covariance matrix $C_{ij}$. However, it is 
diagonal in the basis of the covariance matrix' eigenvectors which makes this basis particular 
convenient for the computation of the PDF error. The vector of the parameters ${P_i}$ 
transformed into the eigenvector basis reads
\begin{equation}
{\tilde {P}_i}=\sum_{k=1}^{N_p} \left(\sqrt{C}\right)^{-1}_{ik}{{P}_k}~,
\label{eq:delp}
\end{equation}
where 
\begin{equation}
\sqrt{C}_{ij}=\sum_{k=1}^{N_p}A_{ik}\sqrt{D}_{kj}~.
\label{eq:covsqrt}
\end{equation}
Here $A_{ik}$ denotes the  matrix with the columns given by the orthonormal eigenvectors 
of $C_{ij}$, $\sqrt{D}_{kj}=\delta_{jk}\sqrt{e_k}$, $e_k$ are the eigenvalues of 
$C_{ij}$, and $\delta_{jk}$ is the Kronecker symbol. 
The PDF uncertainties are commonly presented as the shifts in $\vec q$
due to variation of the parameters ${\tilde P_i}$ by their standard 
deviation. Since the latter are equal to one the shifts are given by
\begin{equation}
\frac{d \vec q}{d{\tilde P_i}}=\sum_{k=1}^{N_p}\frac{d \vec q}{d{P_k}}
\left(\sqrt{C}\right)_{ik}~.
\label{eq:deriv}
\end{equation}
Moreover, the parameters ${\tilde P_i}$ are uncorrelated. Therefore the shifts 
in Eq.~(\ref{eq:deriv}) can be combined in quadrature to obtain the total PDF uncertainty. 
In a similar way the uncertainty in a theoretical prediction $t(\vec {q})$ 
due to the PDFs can be obtained assuming its linear dependence on the PDFs as a combination of 
the variations
\begin{equation}
\Delta t^{(k)}=t\left[\vec q(P_k^0) +\frac{d \vec q}{d{\tilde
      P_k}}\right] - t\left[\vec q(P_k^0)\right]
\label{eq:csvar}
\end{equation}
in quadrature.

Now we show how new data on the hadronic hard-scattering process can be consistently  
accommodated into the PDF fit avoiding involved cross section computations.  Let $P_i^{\rm fit}$ 
the current values of the PDF parameters in the fit with the new data set included and 
$\delta P_i=P_i^{\rm fit}-P_i^{0}$, where $P_i^{0}$ stands for the PDF parameter values 
obtained in the earlier version of the fit performed without the new data-set. The current 
PDF value can be expressed in terms of $\delta P_i$ and the PDF variation in the 
eigenvector basis as follows
\begin{equation}
\vec q ^{\rm fit} = \vec q (P_i^0) + \frac{d \vec q}{d{\tilde P_i}} 
\delta \tilde P_i~,
\end{equation}
where 
\begin{equation}
\delta \tilde P_i=\sum_{k=1}^{N_p} \left(\sqrt{C}\right)^{-1}_{ik}
\delta P_i~.
\end{equation}
A shift in the hard-scattering cross section corresponding to the variation of 
the $i$-th PDF parameter in the fit reads
\begin{equation}
\delta t^{(k)}=t\left[\vec q (P_k^0) + \frac{d \vec
    q}{d{\tilde P_k}} \delta \tilde P_k\right] 
- t\left[\vec q(P_k^0)\right] 
\approx \Delta t^{(k)} \sum_{l=1}^{N_p} \left(\sqrt{C}\right)^{-1}_{il}
\delta P_i
\label{eq:appr}
\end{equation}
and the total change in $t$ is the sum of terms in Eq.~(\ref{eq:appr}) over all parameters 
being fitted. The approximation Eq.~(\ref{eq:appr}) allows fast calculations of the cross section 
for 
the new data added to the PDF fit since the values of $\sigma\left[\vec q(P_i^0)\right]$ and 
$\Delta \sigma_i$ can be prepared in advance. This approach is justified if the variation of 
the parameters in the new fit is localized within their uncertainties obtained in the previous fit 
or in case of sufficient linearity of the PDFs with respect to the fitted parameters and the 
cross sections depending on the PDFs. Furthermore, if the algorithm does not seem to guarantee 
sufficient 
accuracy, it can be applied iteratively, with the update of the  
$\sigma\left[\vec q(P_i^0)\right]$ and $\Delta t_i$ values at 
each iteration. 

{\footnotesize
%

}
\end{document}